\documentclass[structabstract]{aa}
\usepackage[latin1]{inputenc}
\usepackage{txfonts}
\usepackage[pdftex]{graphicx} 
\usepackage[pdftex, plainpages=false]{hyperref}
\usepackage{natbib}
\bibpunct{(}{)}{;}{a}{}{,}

\def\kms {{\mathrm{km}\,\mathrm{s}^{-1}}}

\def\teff {{T_{\mathrm{eff}}}}

\defcitealias{Pereira2009b}{Paper~I}
\defcitealias{CAP2001}{ALA01}
\defcitealias{CAP2004}{AAF04}

\begin{document}

\title{Oxygen lines in solar granulation}
\subtitle{II. Centre-to-limb variation, NLTE line formation, blends and the solar oxygen abundance}
\titlerunning{Oxygen lines in solar granulation. II.}

\author{T. M. D. Pereira\inst{1,2}, M. Asplund\inst{3}, D. Kiselman\inst{2}}
\authorrunning{Pereira et al.}
\institute{Research School of Astronomy and Astrophysics, Australian National University, Cotter Rd., Weston, ACT 2611, Australia\\
\email{tiago@mso.anu.edu.au}
\and The Institute for Solar Physics of the Royal Swedish Academy of Sciences, AlbaNova University Center, 106 91 Stockholm, Sweden
\and Max-Planck-Institut f\"ur Astrophysik, Postfach 1317, D--85741 Garching b. M\"unchen, Germany
}

\date{Received 5 July 2009 / Accepted 9 September 2009}

\abstract 
{There is a lively debate about the solar oxygen abundance and the role of 3D models in its recent downward revision. These models have been tested using high resolution solar atlases of flux and disk-centre intensity. Further testing can be done using centre-to-limb variations.} 
{Using high resolution and high S/N observations of neutral oxygen lines across the solar surface we seek to test if the 3D and 1D models reproduce their observed centre-to-limb variation. In particular we seek to assess whether the latest generation of 3D hydrodynamical solar model atmospheres and NLTE line formation calculations are appropriate to derive the solar oxygen abundance.} 
{We use our recent observations of O\,\textsc{i} 777~nm, O\,\textsc{i} 615.81~nm, [O\,\textsc{i}] 630.03~nm and nine lines of other elements for five viewing angles $0.2\leq\mu\leq 1$ of the quiet solar disk. We compare them with the predicted line profiles from the 3D and 1D models computed using the most up-to-date line formation codes, line data and allowing for departures of LTE. The centre-to-limb variation of the O\,\textsc{i} 777~nm lines is also used to obtain an empirical correction for the poorly known efficiency of the inelastic collisions with  H\,\textsc{i}.} 
{The 3D model generally reproduces the centre-to-limb observations of the lines very well, particularly the oxygen lines. From the O\,\textsc{i} 777~nm lines we find that the classical Drawin recipe slightly overestimates H\,\textsc{i} collisions (we find the best agreement with $S_\mathrm{H}\approx 0.85$). The limb observations of the O\,\textsc{i} 615.82~nm line allow us to identify a previously unknown contribution of molecules for this line, prevalent at the solar limb. A detailed treatment of the [O\,\textsc{i}] 630.03~nm line including the recent nickel abundance shows that the 3D modeling provides an excellent agreement with the observations. The derived oxygen abundances with the 3D model %
are 8.68 (777~nm lines), 8.66 (630.03~nm line) and 8.62 (615.82~nm line).} 
{These additional tests have reinforced the trustworthiness of the 3D model and line formation for abundance analyses.}


\maketitle

\section{Introduction\label{sec:intro}}

Oxygen is arguably one of the most important elements in the Universe. After hydrogen and helium it is the most common element, projecting its importance to many fields in astrophysics: from stellar abundances to inter stellar medium and galactic evolution. The solar oxygen abundance is often used as a reference in many studies. Being a volatile element, its meteoritic abundance is not representative. Hence the most reliable probe for the reference oxygen abundance is the solar photosphere. 

With high-quality data readily available, a long standing expertise in solar observations and much improved knowledge of stellar atmospheres in the last decades, one would expect that a measurement as fundamental as the solar photospheric abundance of oxygen would be firmly established. But, on the contrary, the solar oxygen abundance has been hotly debated in recent times. From the high (and widely used) value of $\log\epsilon_{\mathrm{O}}=8.93\pm0.04$ \citep{AndersGrevesse1989}, the proposed oxygen abundance has been revised downward to the low value of $\log\epsilon_{\mathrm{O}}=8.66\pm0.05$ \citep{Asplund2004}. The pivotal causes for this change have been the proper treatment of the statistical inhomogeneities caused by the solar granulation (making use of a 3D photosphere model); the proper treatment of departures from local thermodynamical equilibrium (LTE) and improved line data (including better blend identification). A low oxygen abundance has caused significant grievances among the solar/stellar interior modelling community by undoing the almost perfect agreement between the solar interior models and helioseismology, which has sparked much debate.  Over the last years the downward revision of the solar oxygen has been supported by some studies \citep[\emph{e.g.}][]{SocasNavarroNorton2007,MelendezAsplund2008} but also contested by others \citep[\emph{e.g.}][]{CentenoSocasNavarro2008,Ayres2008}. Some recent studies find intermediate oxygen abundances \citep[\emph{e.g.}][]{Caffau2008}.

A measure of the photospheric oxygen abundance is made difficult for several reasons. Of the few atomic oxygen lines available in the solar spectrum, some are very weak, others are significantly blended. An example is the popular [O\,\textsc{i}] 630.03~nm line, which is weak and has a non-negligible blend with a nickel line. On the other hand, departures from LTE are also important for some O\,\textsc{i} lines -- the O\,\textsc{i} 777~nm triplet lines being the typical example. These lines are known to show significant departures from LTE, requiring the use of detailed NLTE line formation that requires detailed atomic input data which may not exist.

For the NLTE modeling the collisions with neutral hydrogen are particularly relevant. There is disagreement in the literature regarding the importance of inelastic collisions with H\,\textsc{i}. At this point, the lack of experimental data or quantum mechanical calculations makes its estimation complicated. A customary approach is to adopt a generalization of the classical Drawin formula \citep{Drawin1968}, using the recipe of \citet{Steenbock1984}, often scaled by an empirical factor $S_\mathrm{H}$; note that \citet{Lambert1993} have corrected a mistake in these original formul\ae. However, there are different views regarding which scaling factor to use for oxygen. \citet{Nissen2002} and \citet{Asplund2004} chose $S_\mathrm{H}=0$, based on evidence that the Drawin formula overestimates the H\,\textsc{i} collision efficiency with some atoms for which experimental or detailed quantum mechanical data exists \citep[\emph{e.g.}][]{Barklem_etal2003}. \citet{Caffau2008}, on the other hand, have adopted a seemingly \emph{ad-hoc} value of $S_\mathrm{H}=1/3$.

For the solar O\,\textsc{i} 777~nm lines, the adopted recipe for H\,\textsc{i} collisions has a significant effect on the line shape and strength and consequently, on the derived oxygen abundance. Using the same 3D model of \citet{Asplund2004}, \citet{CAP2004} empirically found that $S_\mathrm{H}=1$ agreed somewhat better than $S_\mathrm{H}=0$ while the LTE case could be ruled out. %

In addition to the line formation physics, the centre-to-limb variation of the lines provides a robust test of the model atmospheres by probing the depth variation of the source functions, as seen from the Eddington-Barbier approximation. In this work we study atomic oxygen lines using observations of their centre-to-limb variation. We study the effects of using different atmospheric models, departures from LTE and properly accounting for blends. Our aim is to provide additional observational tests of some of the models used to infer the oxygen abundance, using new solar observations.

We have obtained high spatial and spectral resolution observations of oxygen and other lines across the solar surface. 
Similar observations of oxygen lines have been obtained in the past \citep[notably][]{Mueller1968,Altrock1968,CAP2004}. However, available data from early works is limited to equivalent widths and the more recent work of \citet{CAP2004} covers only the O\,\textsc{i} 777~nm triplet lines. Furthermore, the long slit used by \citet{CAP2004} increases the uncertainty of the $\mu$ value for their limb data. The observations detailed in the present work include five neutral oxygen lines: the O\,\textsc{i} 615.81~nm, [O\,\textsc{i}] 630.03~nm and the three O\,\textsc{i} 777~nm over five positions in the solar disk. In addition, lines from other elements are also included in the observed spectral regions. %

This paper is organized in the following manner. Our observations are briefly outlined in the next section. In Section~\ref{sec:synth} we detail the model atmospheres and the line formation codes used, which are compared with the observations in Section~\ref{sec:analysis}. Conclusions are made in Section~\ref{sec:disc}.

\section{Observations}

\subsection{Overview}

We make use of the observations of \citet[][hereafter \citetalias{Pereira2009b}]{Pereira2009b}, which we refer to for a detailed description of the programme, instruments and reduction. Whereas \citetalias{Pereira2009b} is focused on high spatial resolution, here we study the centre-to-limb variations of the spatial and temporaly mean spectra. Thus, in terms of the observations the main difference is that we average the spectrograms in space and time to obtain a mean spectrum for each position in the solar disk. 

\begin{figure}
\begin{center}
  \includegraphics[width=0.45\textwidth]{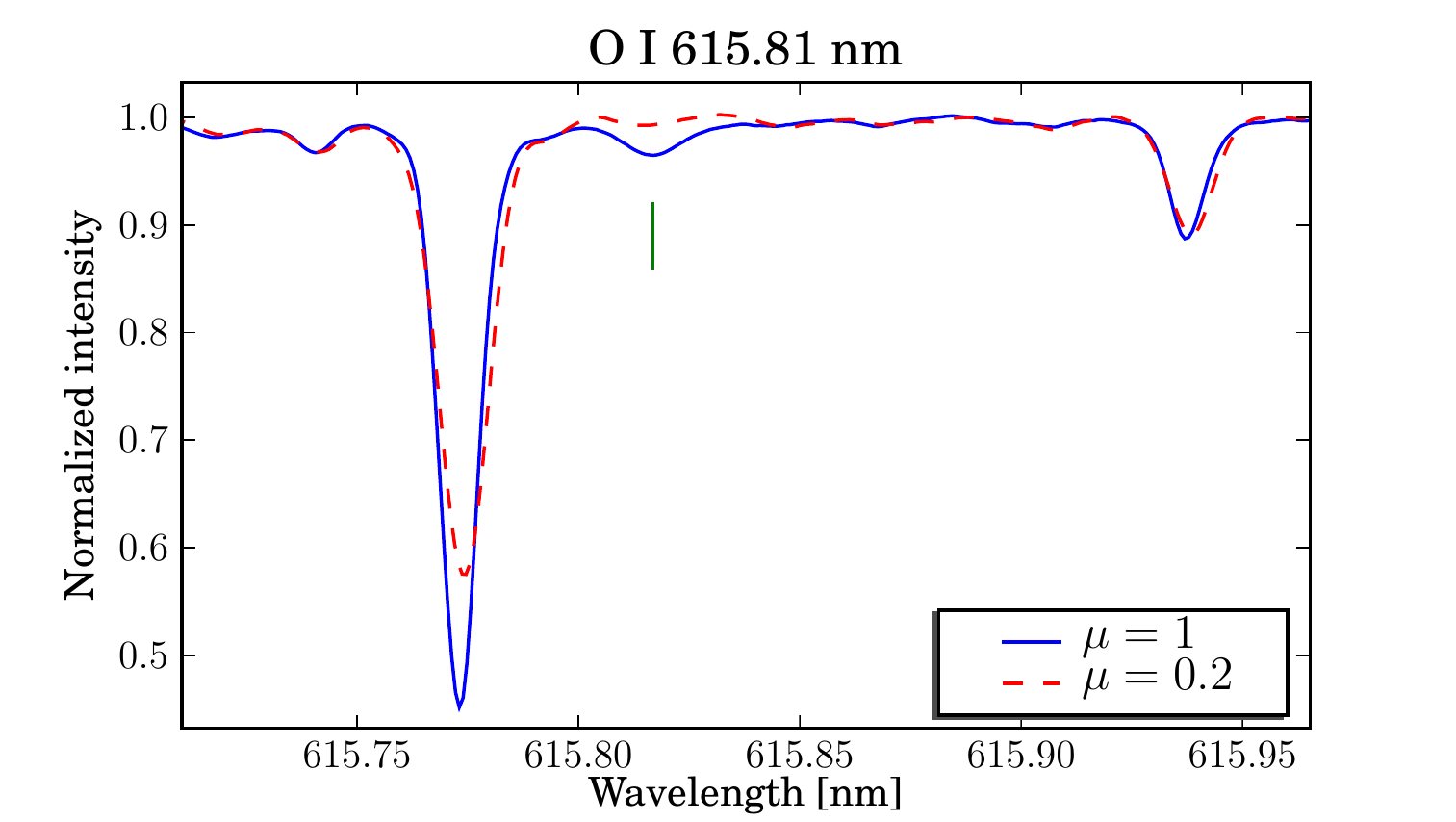}
  \includegraphics[width=0.45\textwidth]{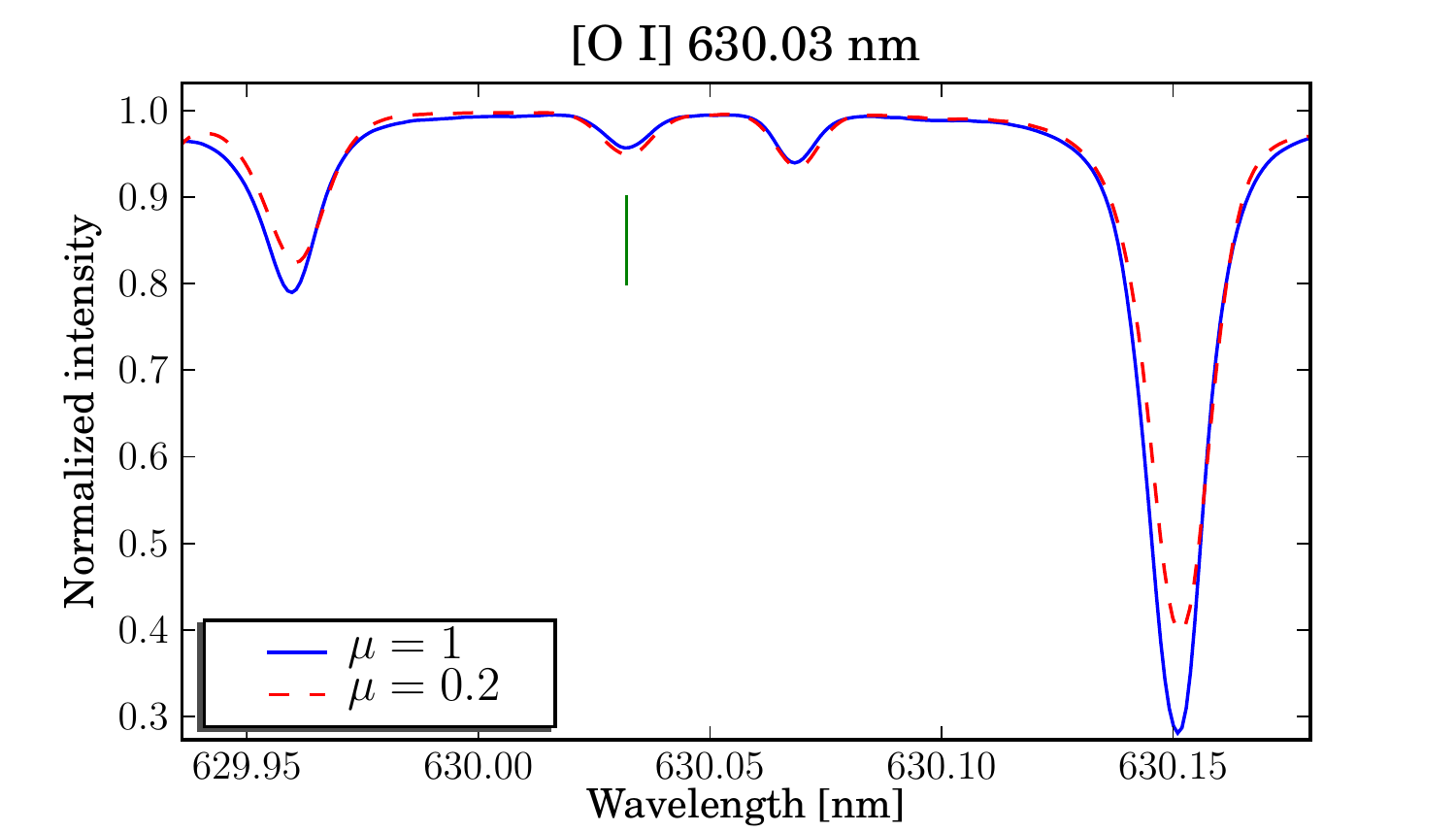}
  \includegraphics[width=0.45\textwidth]{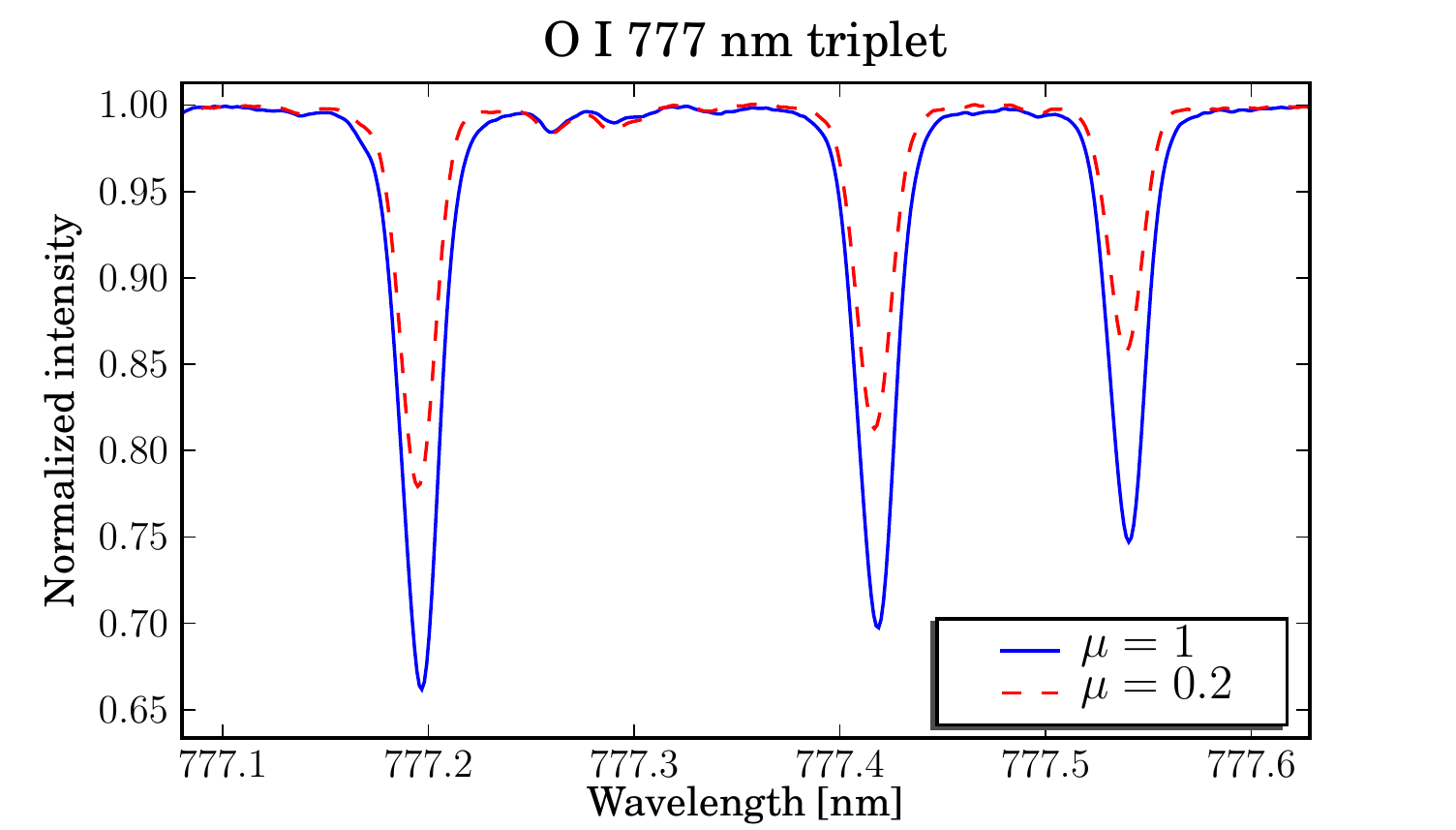}
\end{center}
\caption{Observed profiles at disk-centre and limb. For the weak 615.81 and 630.03~nm lines the location is indicated. In this figure the wavelengths for the limb spectra have been adjusted to compensate for solar rotation and other systematics.}
  \label{fig:obs_profs}
\end{figure}

Using 25 spectrograms for each $\mu\neq 1$ position and 50 spectrograms for $\mu=1$ means that our mean spectra comprise an average of more than 20\,000 spectra ($>$ 40\,000 for disk-centre). The S/N ratio for the spectrum of one spatially averaged spectrogram is about 700. The total S/N would be $\sqrt{25}$ times that value, if the images were all independent -- which is not the case because the time separation does not always guarantee they sample a different granulation pattern. In any case, a conservative estimate of the S/N should be at least 1500 -- more than enough for the present analysis. Unlike in \citetalias{Pereira2009b}, no Fourier filtering is applied, since most of the photon noise is eliminated by the averaging of many spectra.

It is worth to note that for the $\mu\approx 0.4$ set some facular features were visible on the slit-jaw images, though the slit did not cross them. Because we want to study quiet regions, and to avoid any possible interference due to magnetic fields, we have removed the middle third of the spectral images for the $\mu\approx 0.4$ set, so that the final spectrum was averaged from quiet regions only. This procedure reduces the S/N for this set by about 20\%.

\begin{figure*}
\begin{center}
  \includegraphics[width=0.33\textwidth]{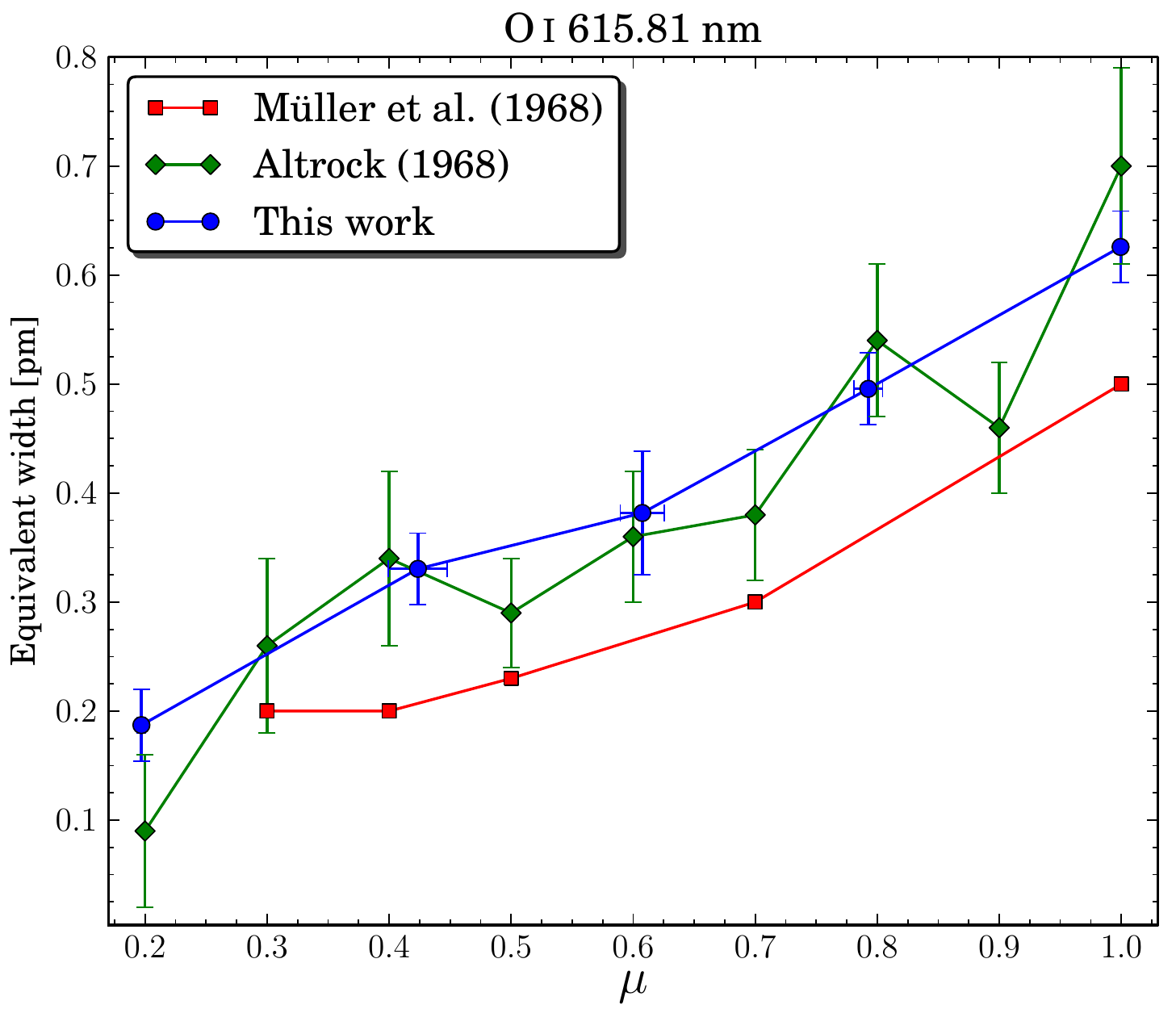}
  \includegraphics[width=0.33\textwidth]{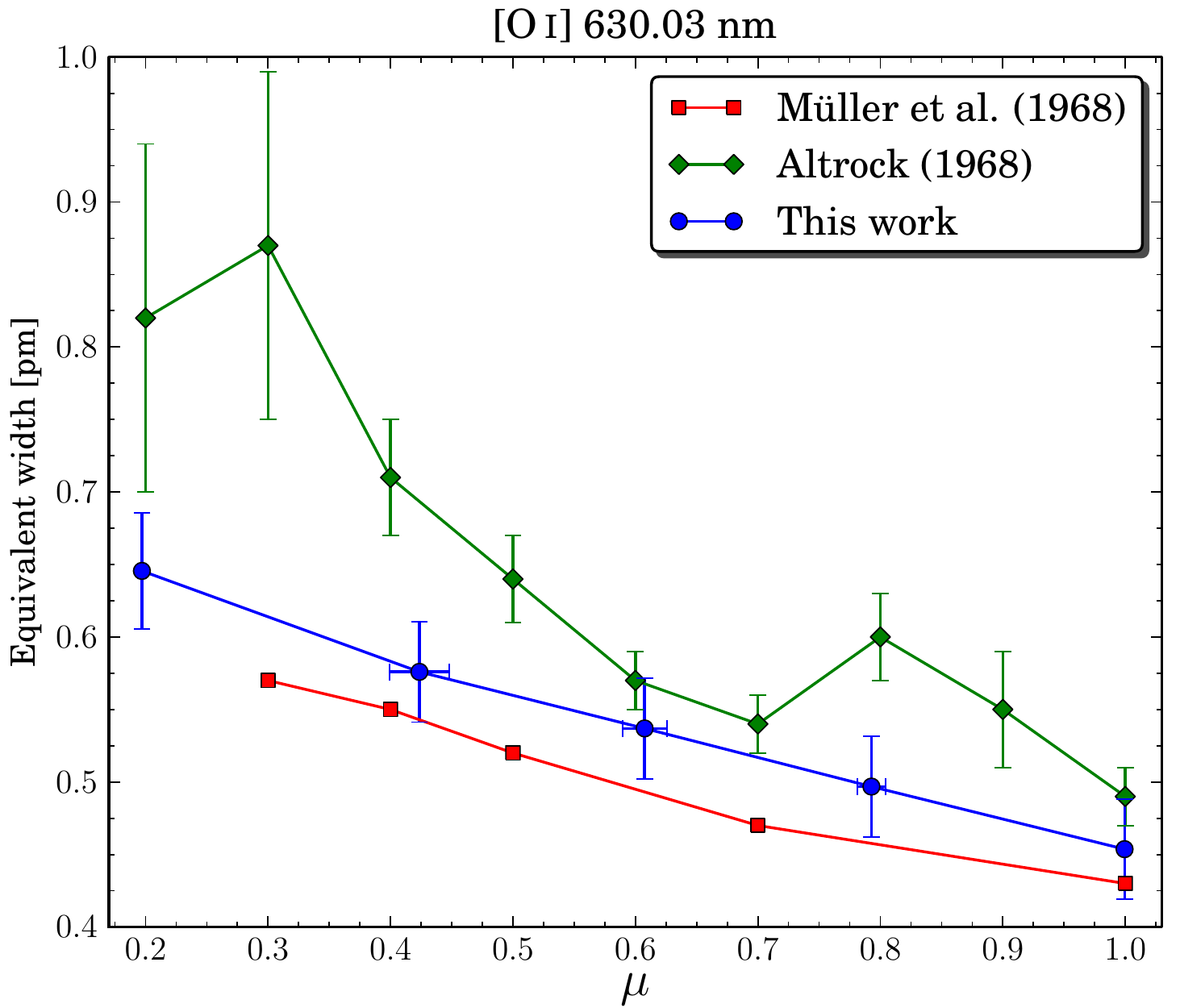}
  \includegraphics[width=0.33\textwidth]{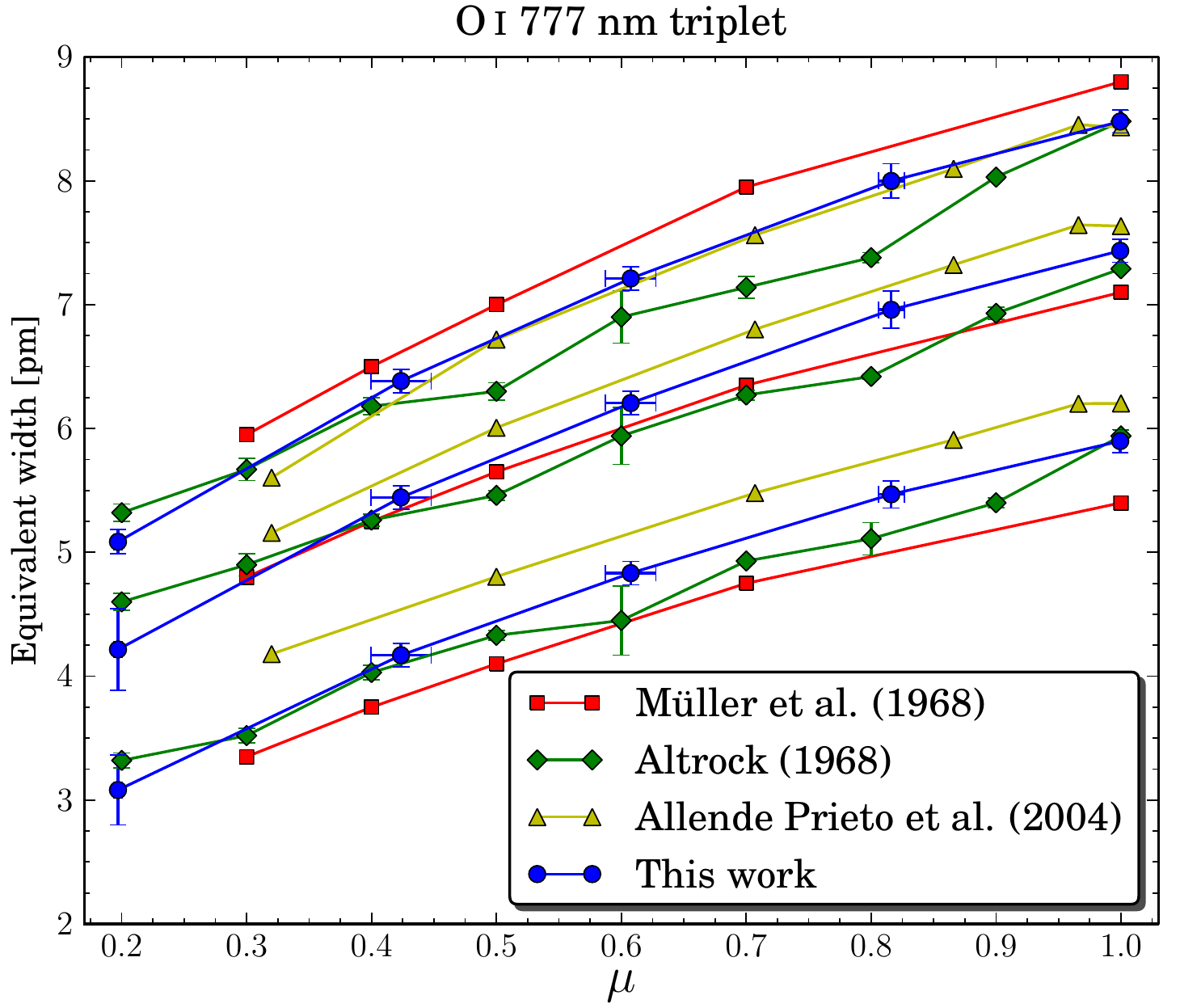}
\end{center}
\caption{Equivalent width vs. $\mu$ for the oxygen lines, for different observations. Last panel shows the three 777~nm O\,\textsc{i} triplet lines (from strongest to weakest: 777.19~nm, 777.41~nm, 777.53~nm). The smooth variation of the equivalent widths with $\mu$ in our data indicates a low statistical error and highlights its high quality. Equivalent widths for \citet{CAP2004} were computed with our analysis routines using their spectra.}
  \label{fig:obs_eqw}
\end{figure*}

After obtaining the spatial and temporally averaged spectra for each $\mu$ the continuum levels are found manually. In some cases like the O\,\textsc{i} 615.81 nm line, a local continuum is used to compensate for blends not included in the line synthesis. The wavelength calibration, detailed in \citetalias{Pereira2009b}, is linked to the Fourier Transform Spectrograph (FTS) disk-centre intensity atlas of \citet{BraultNeckelFTS}. It consists on identifying the same spectral lines in the FTS atlas and our observations, and then obtaining the dispersion relation by a polynomial fit to the line cores. It lacks precision, and for the analysis of some lines a more precise wavelength calibration is necessary. In the analysis section we detail a correction to the wavelength calibration using the atmosphere models and Fe\,\textsc{i} lines. The reduced spectra for disk-centre and the limb can be found in Fig.~\ref{fig:obs_profs}.

\subsection{Comparison with previous studies}

The oxygen lines we observed have been the object of several studies in the past. Most notably, the centre-to-limb variation of these lines has been studied in \citet{Mueller1968}, \citet{Altrock1968} and \citet{CAP2004}. The last work covers only the 777~nm triplet lines. These works use different instrumentation and techniques, which makes them fairly independent of our observations. We compare the equivalent widths of the present work with the ones from these studies in Fig.~\ref{fig:obs_eqw}. Equivalent width measurements have a somewhat subjective component due to different normalizations and different wavelength ranges where it is computed. However, it is the only comparison that can be made at least with \citet{Mueller1968} and \citet{Altrock1968}. With \citet{CAP2004}, we used the available online spectra and computed the equivalent widths in the same way as for our data (direct integration, same wavelength range). It should be noted that their normalization around the 777.41 and 777.53~nm lines is slightly different from ours, leading to a higher continuum level. This in turn makes the lines stronger when computing the equivalent widths by direct integration. The equivalent widths as a function of $\mu$ for all the lines included in this work are listed in Table~\ref{tab:eqws}. The uncertainties arise mostly from the continuum placement. For the lines in the 777 nm region, the errors are larger because a larger integration region was used.

\begin{table*}[ht]
\caption{Equivalent widths for the lines included in this analysis, as a function of $\mu$.}
\label{tab:eqws}
\centering
\begin{tabular}{ccrrrrr}
\hline\hline
  \multicolumn{2}{c}{Line} & \multicolumn{5}{c}{$W$ [pm]}\\
Species & $\lambda$ [nm] & $\mu=0.999\pm 0.001$ & $\mu=0.793\pm 0.012$ & $\mu=0.608\pm 0.018$ & $\mu=0.424\pm 0.024 $ & $\mu=0.197 \pm 0.003 $ \\
\hline
&&&&&&\\
Fe\,\textsc{i} & 615.1618  & $4.95\pm 0.04$ & $5.14\pm 0.04$ & $5.19\pm 0.04$ & $5.55\pm 0.04$ & $5.65\pm 0.04$ \\
Si\,\textsc{i} & 615.5693  & $0.61\pm 0.03$ & $0.62\pm 0.03$ & $0.68\pm 0.03$ & $0.67\pm 0.03$ & $0.62\pm 0.03$ \\
Ca\,\textsc{i} & 615.6023  & $0.92\pm 0.03$ & $0.92\pm 0.03$ & $1.01\pm 0.08$ & $0.97\pm 0.08$ & $1.14\pm 0.03$ \\
Fe\,\textsc{i} & 615.7728  & $6.45\pm 0.04$ & $6.49\pm 0.04$ & $6.39\pm 0.05$ & $6.58\pm 0.16$ & $6.42\pm 0.04$ \\
O\,\textsc{i}  & 615.8186  & $0.63\pm 0.03$ & $0.50\pm 0.03$ & $0.38\pm 0.06$ & $0.33\pm 0.03$ & $0.19\pm 0.03$ \\
Fe\,\textsc{i} & 615.9378  & $1.21\pm 0.04$ & $1.28\pm 0.07$ & $1.31\pm 0.06$ & $1.39\pm 0.11$ & $1.39\pm 0.04$ \\
Fe\,\textsc{i} & 629.0965  & $7.06\pm 0.05$ & $7.05\pm 0.05$ & $6.92\pm 0.05$ & $6.87\pm 0.15$ & $6.36\pm 0.05$ \\
Fe\,\textsc{i} & 629.7793  & $7.54\pm 0.04$ & $7.53\pm 0.04$ & $7.52\pm 0.15$ & $7.77\pm 0.04$ & $7.71\pm 0.04$ \\
$$[O\,\textsc{i}]&630.0304 & $0.45\pm 0.04$ & $0.50\pm 0.04$ & $0.54\pm 0.04$ & $0.58\pm 0.04$ & $0.65\pm 0.04$ \\
Sc\,\textsc{ii}&630.0698   & $0.59\pm 0.03$ & $0.65\pm 0.03$ & $0.68\pm 0.03$ & $0.76\pm 0.03$ & $0.77\pm 0.03$ \\
O\,\textsc{i} & 777.1941   & $8.48\pm 0.09$ & $8.00\pm 0.14$ & $7.21\pm 0.09$ & $6.38\pm 0.10$ & $5.09\pm 0.10$ \\
O\,\textsc{i} & 777.4166   & $7.44\pm 0.09$ & $6.96\pm 0.15$ & $6.21\pm 0.09$ & $5.44\pm 0.10$ & $4.22\pm 0.33$ \\
O\,\textsc{i} & 777.5390   & $5.90\pm 0.09$ & $5.47\pm 0.11$ & $4.83\pm 0.09$ & $4.17\pm 0.10$ & $3.08\pm 0.28$ \\
Fe\,\textsc{i}& 778.0557   &$14.81\pm 0.14$ &$14.45\pm 0.14$ &$14.13\pm 0.14$ &$13.39\pm 0.14$ &$11.66\pm 0.14$ \\
\hline\hline
\end{tabular}
\end{table*}

\section{Theoretical line profiles\label{sec:synth}}

\begin{figure}
  \centering
  \includegraphics[width=0.49\textwidth]{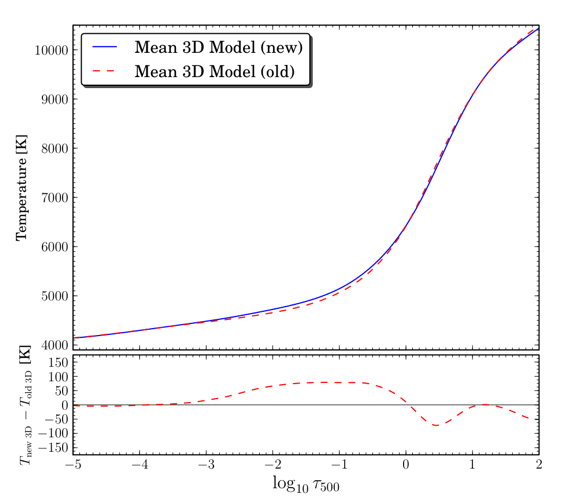}
\caption{Comparison of the mean temperature structure of the new 3D model used in this work and the `old' 3D model of \citet{Asplund2000}, plotted against the optical depth at 500~nm.}
  \label{fig:ttau}
\end{figure}

We employ the 3D time-dependent hydrodynamical simulation of the solar photosphere used in \citetalias{Pereira2009b} (Trampedach et al., in preparation; Asplund et al., in preparation). This model was computed using a more recent version of the Stein \& Nordlund codes. When compared with the 3D model of \citet{Asplund2000}, it includes more detailed radiative transfer: 12 opacity bins, additional sources of continuum opacities such as photo-ionization cross-sections from the Opacity and Iron Projects \citep{Cunto1993,Hummer1993}, etc. In Fig.~\ref{fig:ttau} we show the differences in the mean temperature structure (averaged over surfaces of constant optical depth) between both 3D models.

A comparison is also made with one-dimensional, hydrostatic, time independent solar model atmospheres, including the semi-empirical model of \citet{HM1974} and the \textsc{marcs} model \citep{MARCS2008}.

We computed the synthetic LTE and NLTE (oxygen only) line profiles, making use of our LTE line formation code and the \textsc{multi3d} code \citep{Botnen1997,Asplund2003a}. For a more detailed description of the 3D model, line formation and atomic data used we refer to \citetalias{Pereira2009b}. However, a few things are done differently from \citetalias{Pereira2009b}, as discussed below.

In this work we make use of 90 snapshots of the 3D simulation (in \citetalias{Pereira2009b} we use 20 snapshots). These cover $\approx 45$~min of solar time and have  $\langle\teff\rangle=5778\pm 14\:\mathrm{K}$. To save computational time, 3D NLTE line formation is carried only on four snapshots (chosen to have a representative range of effective temperature and time coverage). Using \textsc{multi3d} we compute NLTE and LTE line profiles, which are used to obtain the wavelength-dependent NLTE/LTE ratio for each $\mu$ value. In contrast with \citetalias{Pereira2009b}, the NLTE/LTE ratio is averaged over all the spatial points and the four snapshots. For each $\mu$ value, we then multiply the NLTE/LTE ratio by the spatially and temporally averaged LTE line profiles (computed with our LTE code for all the 90 snapshots). The NLTE line profiles were computed for eight different values of $S_{\mathrm{H}}$: 0.01, 0.1, 0.3, 0.5, 1, 1.5, 3, 10.

The line profiles from the 1D models, based on the \textsc{marcs} and Holweger--M\"uller model atmospheres, were computed using the same procedures, opacities and radiative transfer. The exception is that a microturbulence of $\xi_{\mathrm{turb}}=1.0\: \kms$ has been used when computing these line profiles. In addition, 1D line profiles were convolved with a Gaussian to account for macroturbulence. The adopted macroturbulences vary from line to line, as detailed in Sect.~\ref{sec:analysis}. For consistency, 1D NLTE line formation is computed in a similar manner to the 3D model, obtaining the NLTE to LTE ratio from \textsc{multi3d} and multiplying it by the profiles obtained with the LTE code (even though for the 1D models there is only one snapshot).

Our main focus are oxygen lines. However, a few lines from other elements are also present in our observations. These lines and their properties, along with the oxygen lines, are listed in Table~\ref{tab:lines}. In Sect.~\ref{sec:other} we briefly discuss the results for the centre-to-limb variation of these lines.

For the oxygen lines and, when available, for lines of other elements the collisional (van der Waals) broadening was computed using the quantum mechanical theory of \citet{AnsteeOMara1995,Barklem1997,Barklem1998}, avoiding the need for conventional \citet{Unsold1955} enhancement factors.

\begin{table}[ht]
\caption{Lines studied in the present work and their parameters.}
\label{tab:lines}
\begin{center}
\begin{tabular}{ccrrr}
\hline\hline
Atomic Species & $\lambda$ [nm] & $\log gf$ & $E_{\mathrm{low}}$ [eV] & $W_{\mu=1}$ [pm]$^{\rm{a}}$ \\
\hline
Fe\,\textsc{i} & 615.1618$^{\rm{d}}$ & -3.299$^{\rm{c}}$ &  2.18$^{\rm{c}}$ & 4.95\\
Si\,\textsc{i} & 615.5693$^{\rm{b}}$ & -2.252$^{\rm{b}}$ &  5.62$^{\rm{b}}$ & 0.61\\
Ca\,\textsc{i} & 615.6023$^{\rm{b}}$ & -2.497$^{\rm{b}}$ &  2.52$^{\rm{b}}$ & 0.92\\
Fe\,\textsc{i} & 615.7728$^{\rm{d}}$ & -1.260$^{\rm{b}}$ &  4.08$^{\rm{c}}$ & 6.45\\
O\,\textsc{i}  & 615.8186$^{\rm{c}}$ & -1.841$^{\rm{c}}$ & 10.74$^{\rm{c}}$ & 0.63\\
Fe\,\textsc{i} & 615.9378$^{\rm{d}}$ & -1.970$^{\rm{b}}$ &  4.61$^{\rm{b}}$ & 1.21\\
\hline
Fe\,\textsc{i} & 629.0965$^{\rm{b}}$ & -0.774$^{\rm{b}}$ & 4.73$^{\rm{b}}$ & 7.06\\
Fe\,\textsc{i} & 629.7792$^{\rm{c}}$ & -2.740$^{\rm{c}}$ & 2.22$^{\rm{c}}$ & 7.54\\
$$ [O\,\textsc{i}] & 630.0304$^{\rm{c}}$ & -9.717$^{\rm{e}}$ & 0.00$^{\rm{c}}$ & 0.45\\
Sc\,\textsc{ii} & 630.0698$^{\rm{b}}$ & -1.887$^{\rm{b}}$ & 1.51$^{\rm{b}}$    & 0.59\\
\hline
O\,\textsc{i}  & 777.1944$^{\rm{c}}$ & 0.369$^{\rm{c}}$ & 9.15$^{\rm{c}}$ & 8.48\\
O\,\textsc{i}  & 777.4166$^{\rm{c}}$ & 0.223$^{\rm{c}}$ & 9.15$^{\rm{c}}$ & 7.44\\
O\,\textsc{i}  & 777.5390$^{\rm{c}}$ & 0.002$^{\rm{c}}$ & 9.15$^{\rm{c}}$ & 5.90\\
Fe\,\textsc{i} & 778.0557$^{\rm{b}}$ & 0.029$^{\rm{b}}$ & 4.47$^{\rm{b}}$  & 14.81\\
\hline\hline
\end{tabular}
\end{center}
\begin{list}{}{}
  \item[$^{\rm{a}}$] Equivalent widths at disk-centre measured from our observations
  \item[$^{\rm{b}}$] VALD database \citep{VALD1,VALD2}
  \item[$^{\rm{c}}$] NIST database \citep{NIST}
  \item[$^{\rm{d}}$] \citet{Nave1994}
  \item[$^{\rm{e}}$] \citet{Storey2000}
\end{list}
\end{table}

\section{Results\label{sec:analysis}}

The synthetic line profiles are compared with the observations by means of fitting and equivalent width. Before a comparison was made with our observations, the synthetic line profiles were convolved with a Gaussian (equivalent to $\lambda/\Delta\lambda$=200\,000) to account for the instrumental profile of the spectrograph. When comparing line profiles with the FTS intensity atlas no convolution took place, due to the very high resolution of the atlas. The line profile fitting was performed using the L-BFGS-B optimization algorithm \citep{l-bfgs-b}%

\begin{figure*}
  \includegraphics[width=0.33\textwidth]{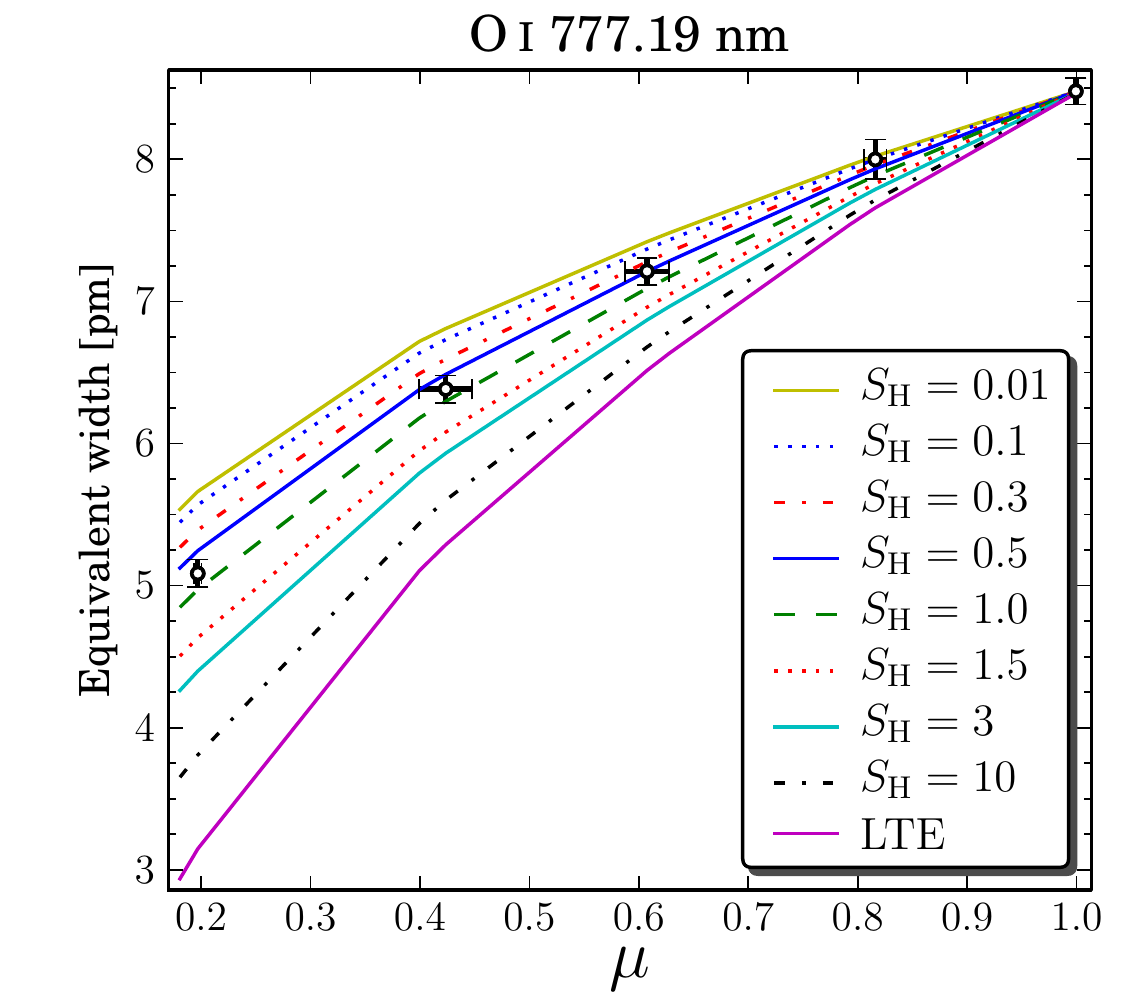}
  \includegraphics[width=0.33\textwidth]{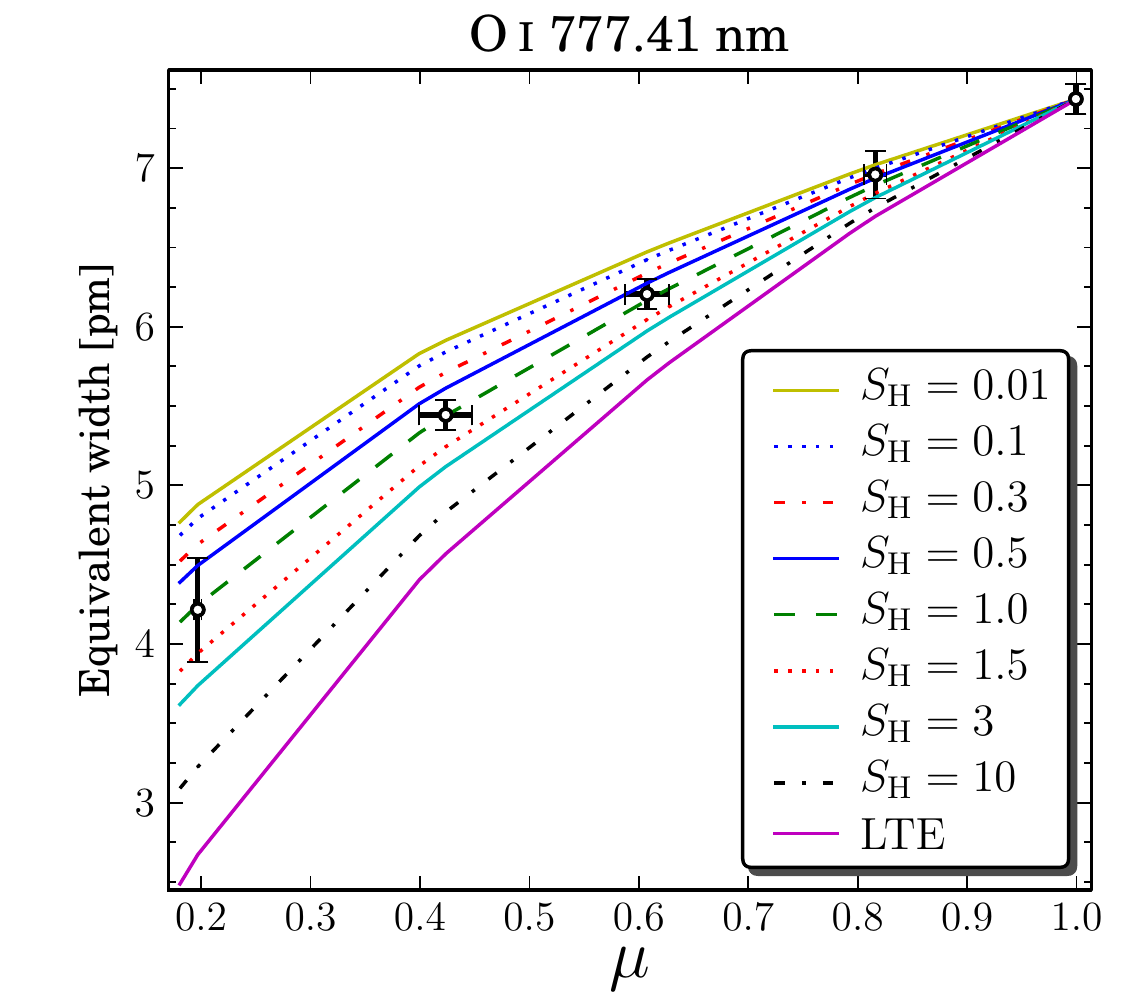}
  \includegraphics[width=0.33\textwidth]{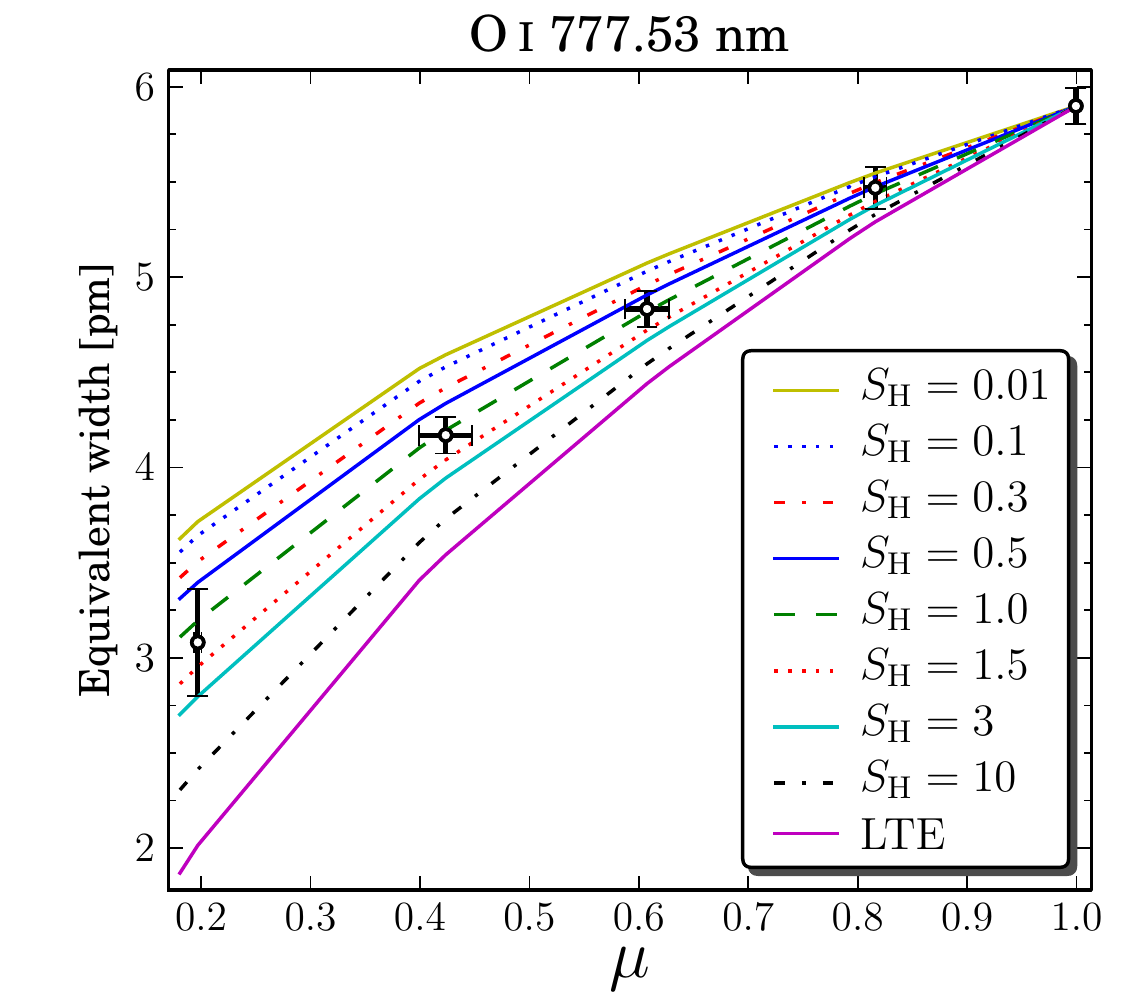}
  \includegraphics[width=0.33\textwidth]{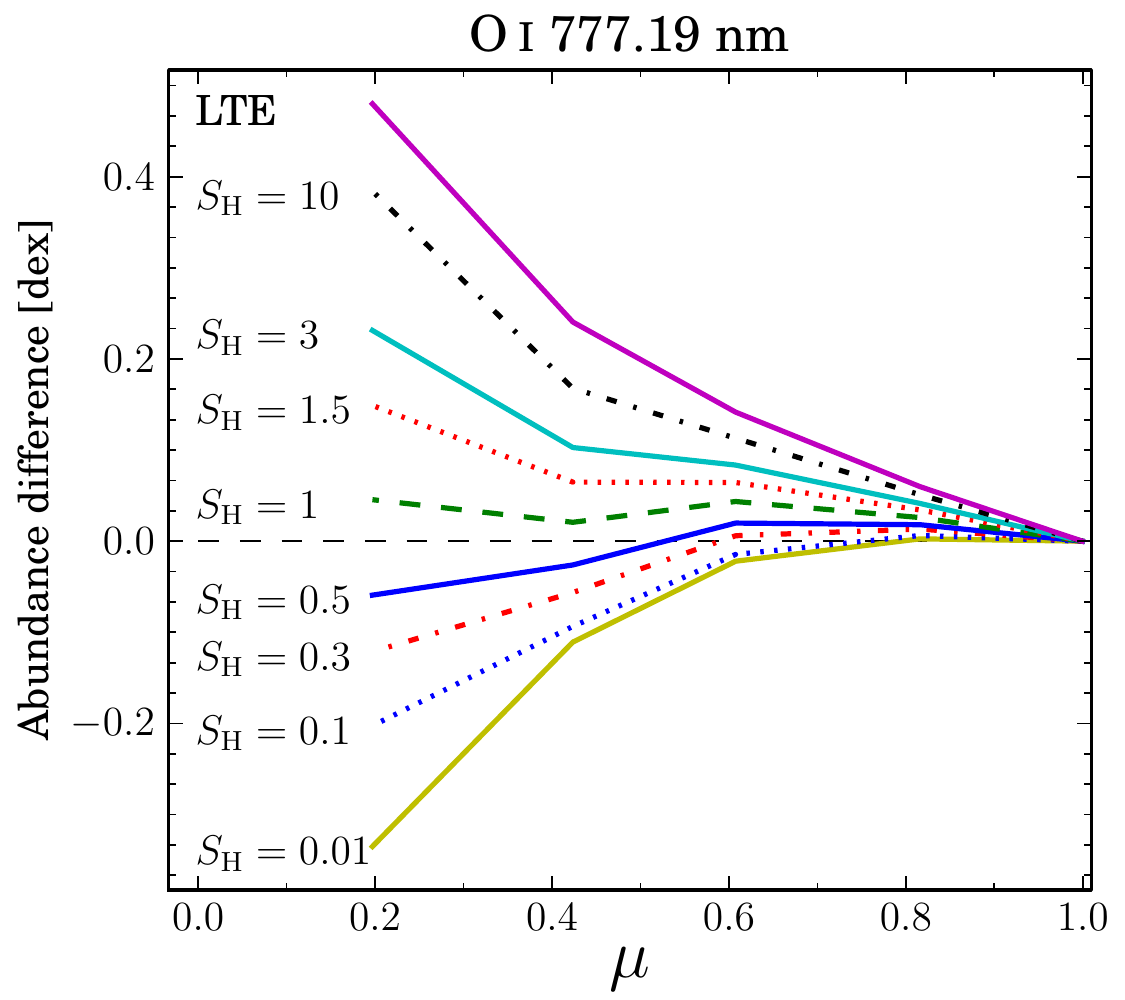}
  \includegraphics[width=0.33\textwidth]{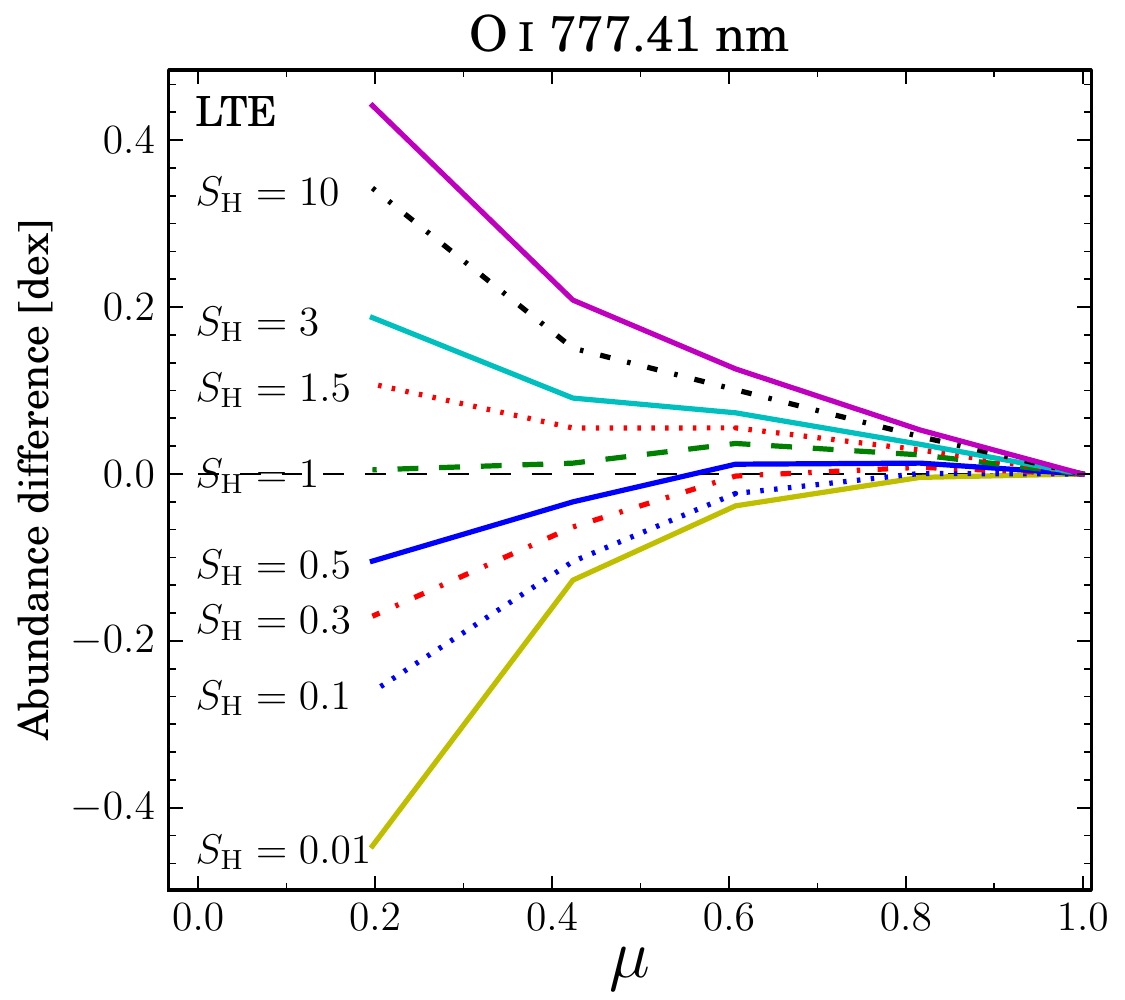}
  \includegraphics[width=0.33\textwidth]{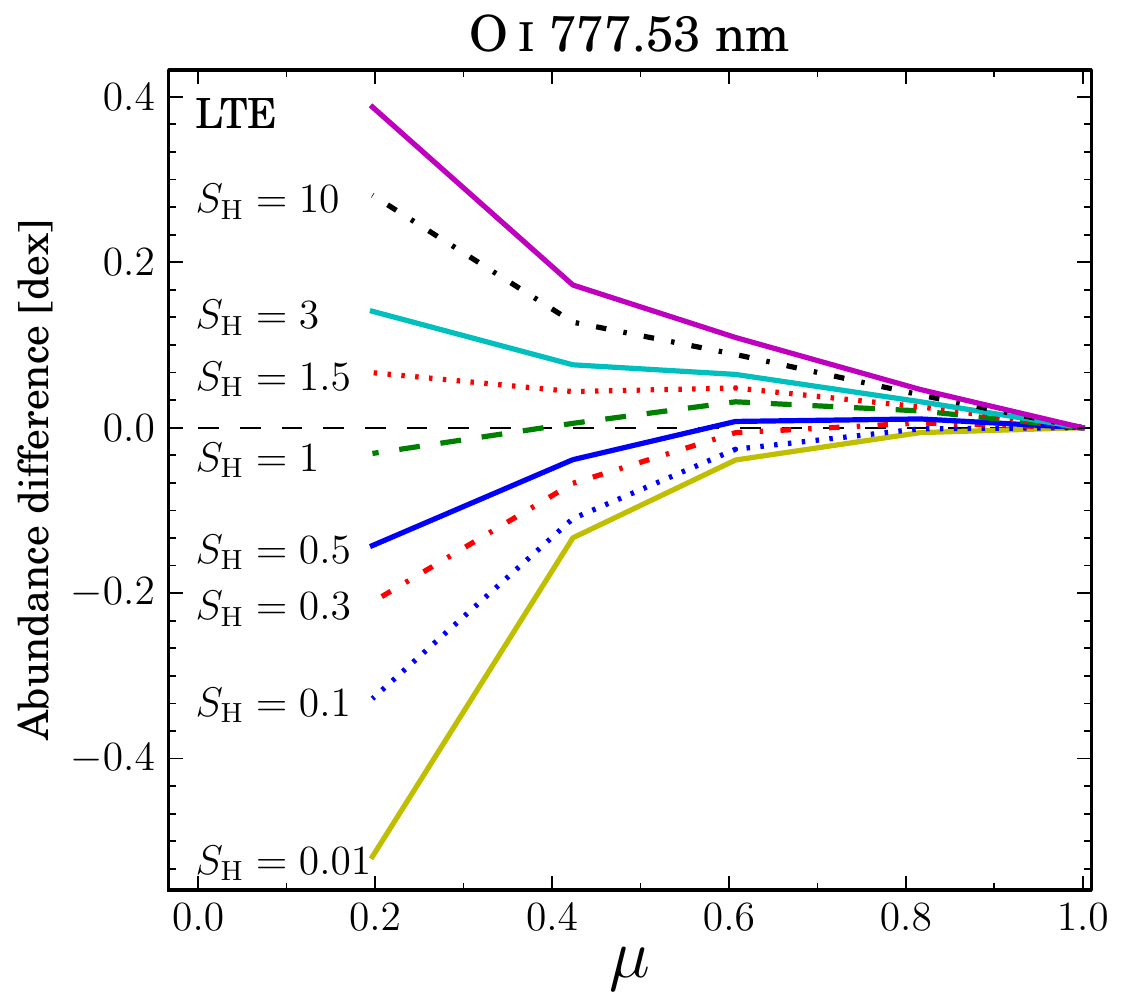}
\caption{Top panels: equivalent width vs. $\mu$ for the three triplet lines, using 3D LTE and NLTE with different $S_{\mathrm{H}}$ coefficients for hydrogen collisions. The oxygen abundance was adjusted so that the models have the same equivalent width as the observations at disk-centre. Bottom panels: difference in fitted abundance from disk-centre to a given position in $\mu$. Synthetic profiles were fitted against the observations at each $\mu$, and the figures show how much the fitted abundance varies from the fitted value at disk-centre.}
  \label{fig:777_results3d}
\end{figure*}

\subsection{O\,I infrared triplet}
\subsubsection{Context}

The three O\,\textsc{i} lines around 777~nm are strong and lie in a part of the solar spectrum relatively free of blends and telluric lines. These two characteristics alone make them a good abundance indicator. However, these lines show significant departures from LTE in the Sun \citep{Altrock1968}, because of a radiation field weaker than Planckian in the line formation region \citep[\emph{e.g.}][]{Eriksson1979,Kiselman1993}. 

Aside from the increased computational expense of computing the 3D NLTE radiative transfer for these lines, departures from LTE introduce additional uncertainties stemming from the input physics, in particular photo-ionization rates and collisional cross-sections with electrons and H\,\textsc{i}. The latter, as noted in Sect.~\ref{sec:intro} are often taken from the classical estimates and scaled by an  $S_\mathrm{H}$ factor. For the O\,\textsc{i} 777~nm lines a scaling factor for the H\,\textsc{i} collisions was tried by \citet{Kiselman1993}, to reconcile different 1D models with observed centre-to-limb variation of the equivalent widths. Further investigation on 3D and NLTE effects on these lines was provided by \citet{Kiselman1995}, using an early 3D model and showing the feasibility of the line centre-to-limb variations to probe for the atomic parameters (i.e., $S_\mathrm{H}$). This suggestion was followed by \citet[hereafter AAF04]{CAP2004}, who used the centre-to-limb variation of the O\,\textsc{i} 777~nm lines and a 3D model to empirically deduce that $S_\mathrm{H}=1$ was preferrable to $S_\mathrm{H}=0$.

Our work differs from \citetalias{CAP2004} in that we use a different set of observations and 3D model, a greater range of $S_\mathrm{H}$ values, a more up-to-date atom and more detailed radiative transfer for the NLTE calculations (more simulation snapshots included in the full 3D NLTE, more angles used in \textsc{multi3d}).

\subsubsection{Comparison with observations}

\begin{figure} 
  \centering
  \includegraphics[width=0.24\textwidth]{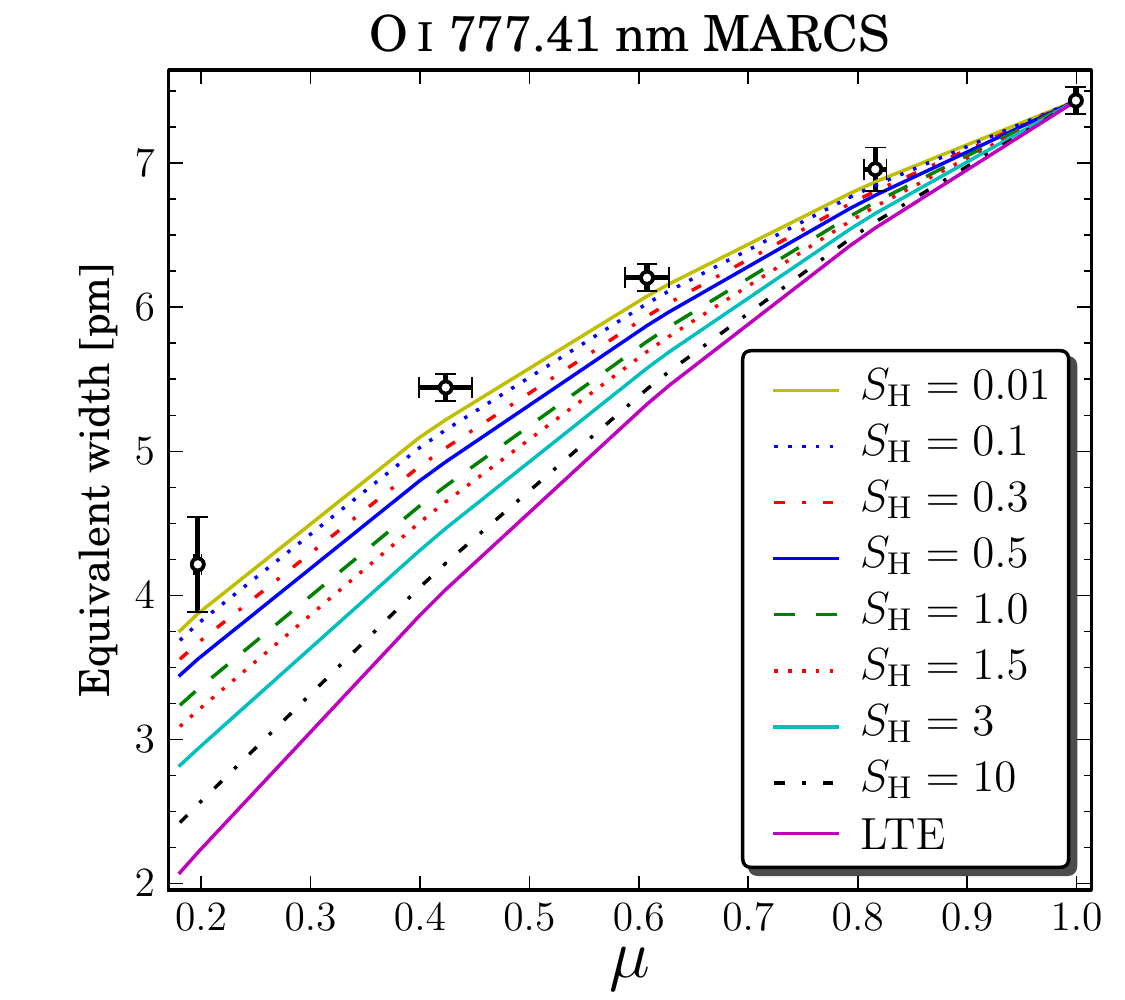}
  \includegraphics[width=0.24\textwidth]{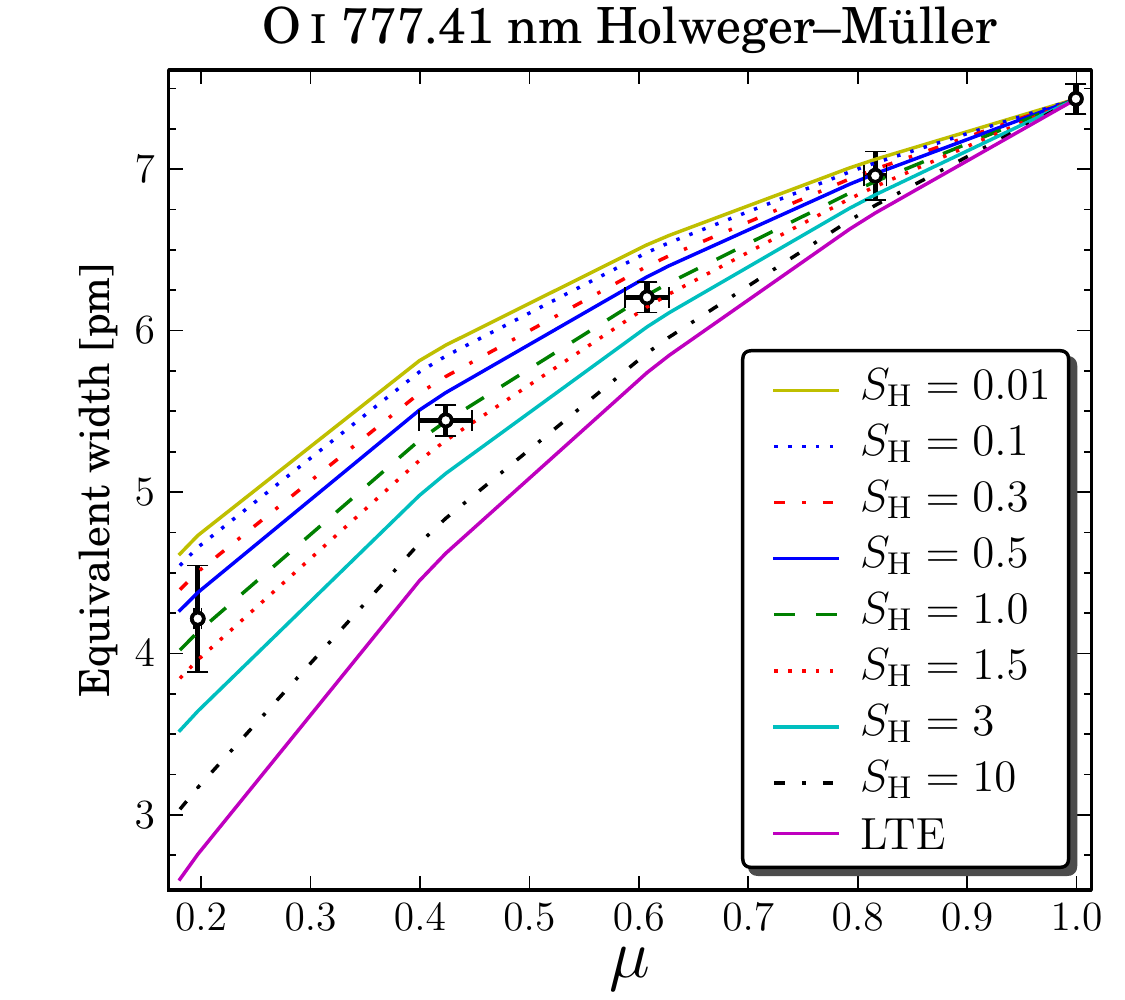}\\
  \includegraphics[width=0.24\textwidth]{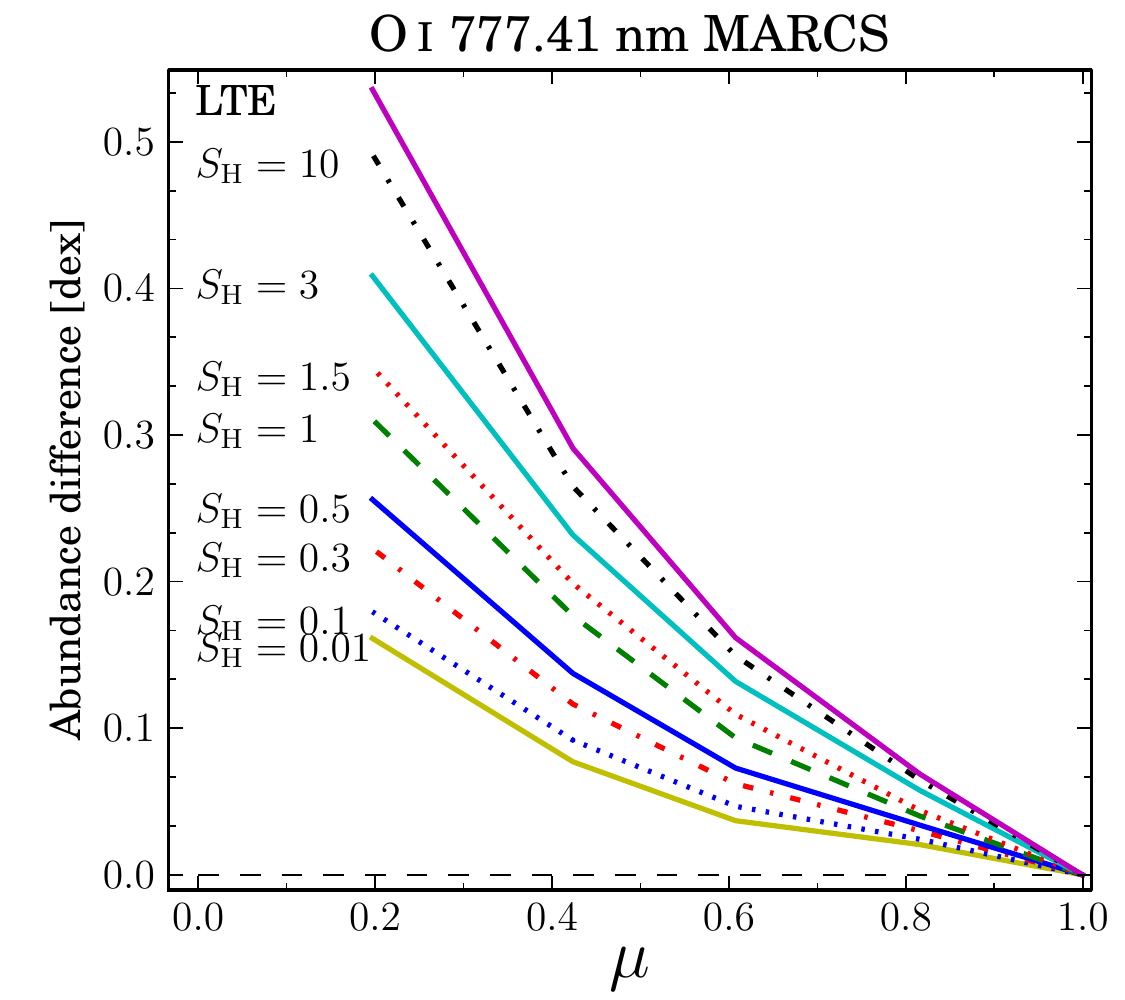}
  \includegraphics[width=0.24\textwidth]{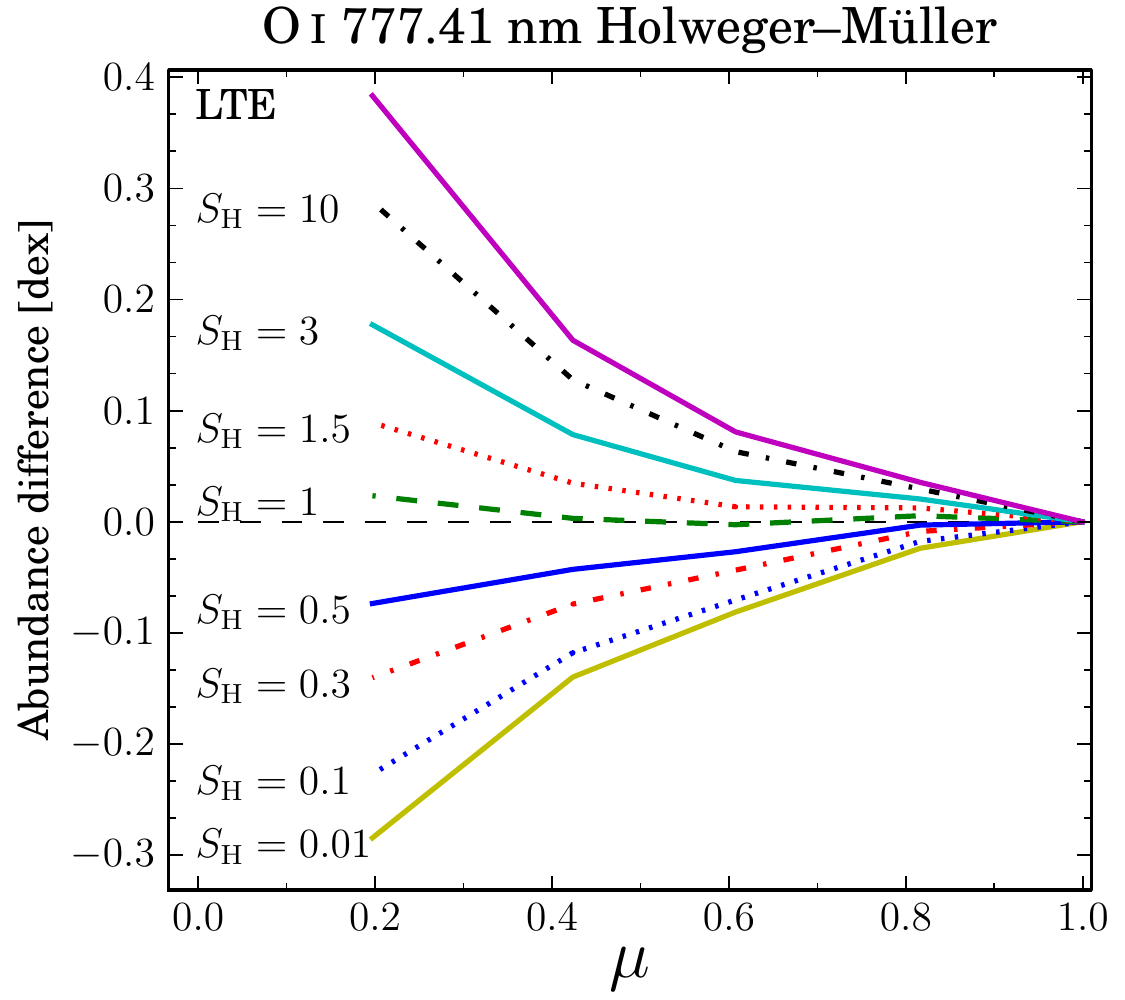}
\caption{Same as Fig.~\ref{fig:777_results3d} but for the 1D models and only for the O\,\textsc{i} \mbox{777.41~nm} line.}
  \label{fig:777_results1d}
\end{figure}

Our goal is to find the $S_\mathrm{H}$ that best describes the centre-to-limb variation of strength of the lines. For this purpose we have two diagnostics: equivalent widths and line profile fitting. 

Measuring the variation of line strengths by profile fitting has the advantage of being less sensitive to systematics (\emph{e.g.} blends, noise) than the equivalent widths. But when the shape of the line profiles does not match the observations, the line profiles adjusted in abundance will have a slightly different equivalent width than the observations. We present the results using both diagnostics, and they both indicate a very similar result.

To fit the line profiles we vary two parameters: oxygen abundance and wavelength shift (of the observations). The latter is necessary because of the uncertainty in the wavelength calibration. For the 1D models another free parameter was allowed in the fit: macroturbulence. Ideally it should not be allowed to vary freely, but rather extracted from nearby lines. The scarcity of lines in our observed window around 777~nm makes this task difficult; in our observed window there is only one other line, a strong Fe\,\textsc{i} line. Deriving the macroturbulence from only one line would probably introduce a similar or larger error as allowing it to be a free parameter in the fit for the O\,\textsc{i} 777~nm lines. %

The results for profile fitting and equivalent width for the 3D model are given in Fig.~\ref{fig:777_results3d}. For the equivalent widths, the oxygen abundance was adjusted for each line so that it matched the observations at disk-centre. The same abundance was then used for all the other values of $\mu$. Both diagnostics indicate that $S_\mathrm{H}=1$ gives the best agreement with the observations, with a small variation from line to line. It is shown with a high confidence level that LTE line formation is not a valid approximation for these lines. In Table~\ref{table:abund77} we list the derived abundances from fitting the disk-centre line profiles. They were fitted in the following wavelength ranges: 777.175--777.24~nm, 777.36--777.46~nm and 777.50--777.565~nm.

Corresponding results for the 1D models are given in Fig.~\ref{fig:777_results1d}, but only for the O\,\textsc{i} 777.41~nm line; the other lines behave similarly. They indicate a mixed scenario. On one hand, the Holweger--M\"{u}ller model gives a similar result to the 3D model: the best agreement is with $S_\mathrm{H}=1$. But for the \textsc{marcs} model no value of $S_\mathrm{H}$ can give a reasonable agreement with the observations, both in equivalent width and line profile fitting. %

\begin{figure*}
  \centering
  \includegraphics[width=0.6\textwidth]{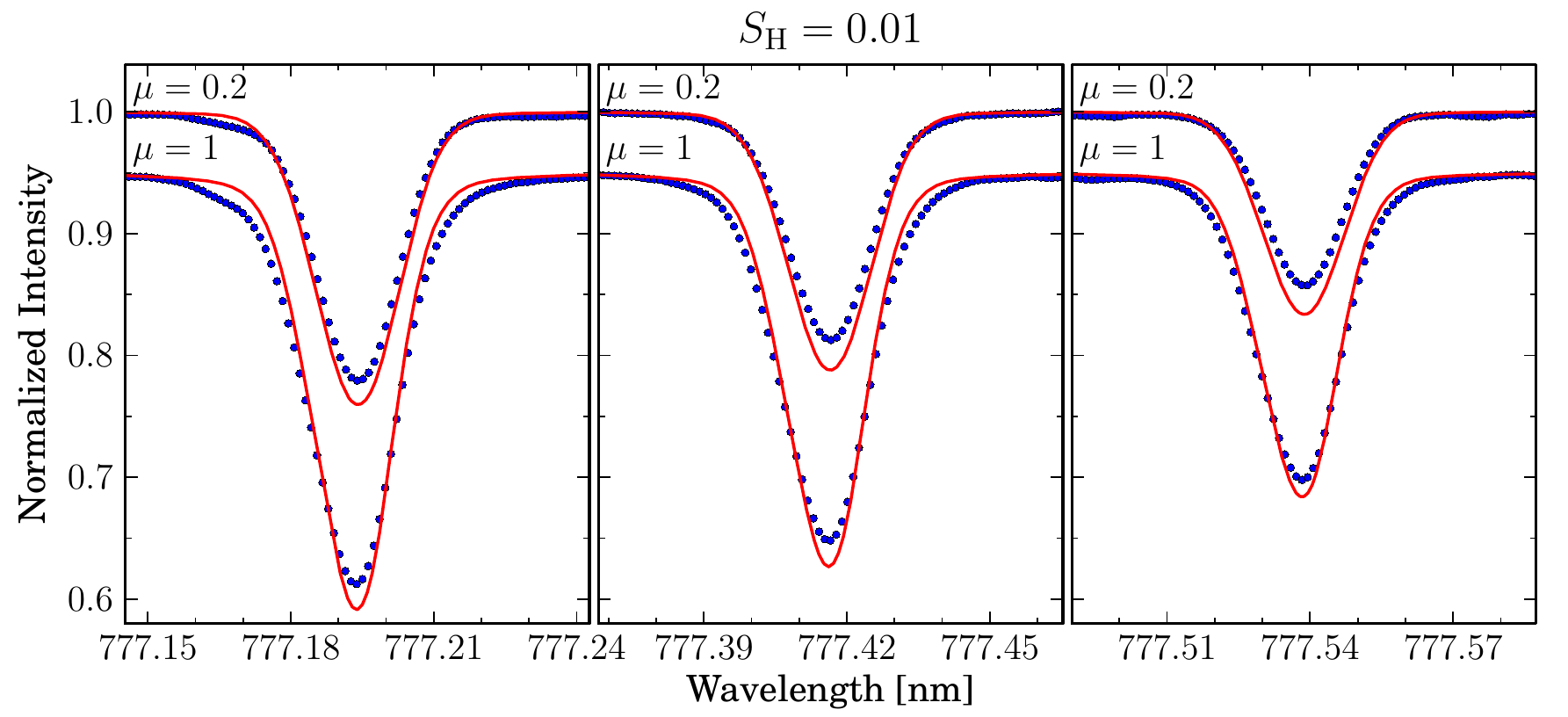} \\
  \includegraphics[width=0.6\textwidth]{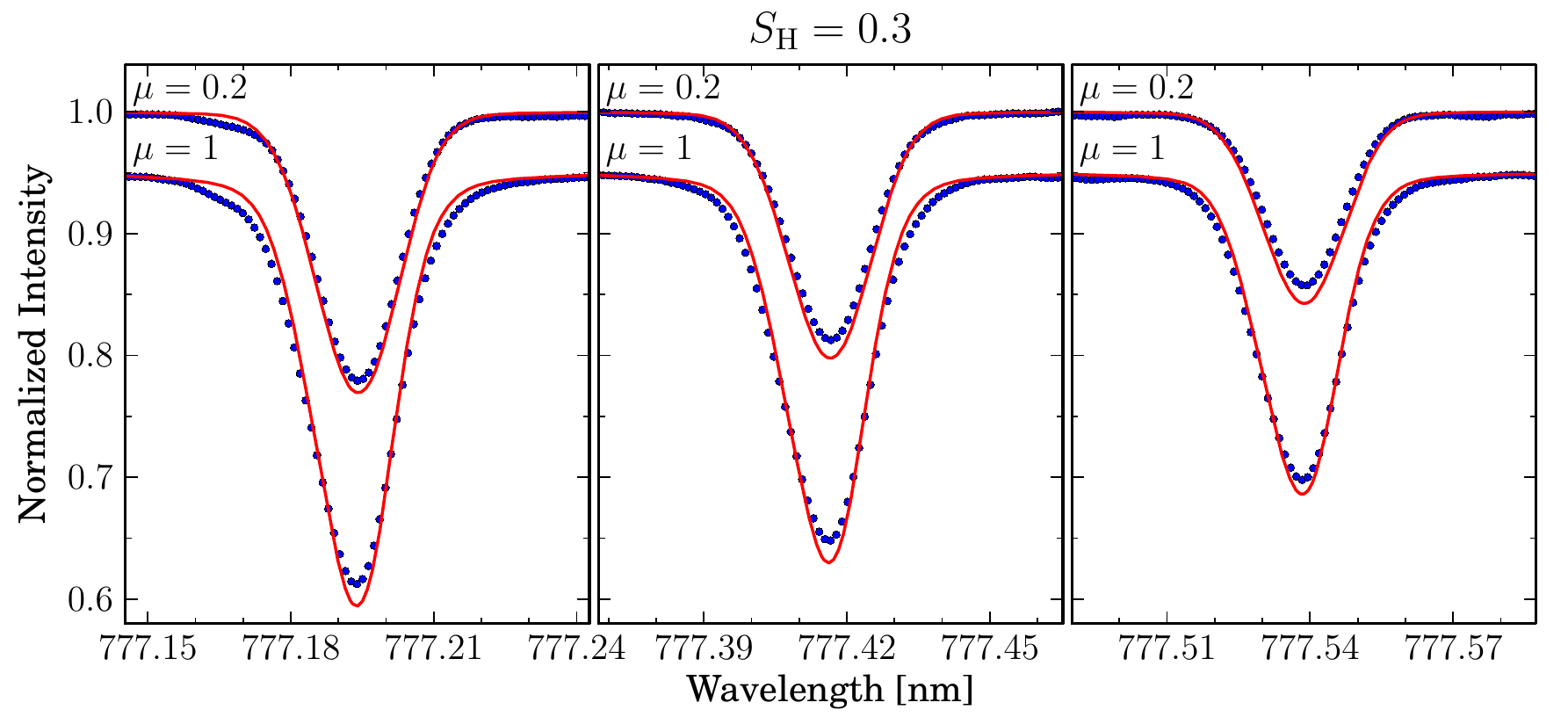}  \\
  \includegraphics[width=0.6\textwidth]{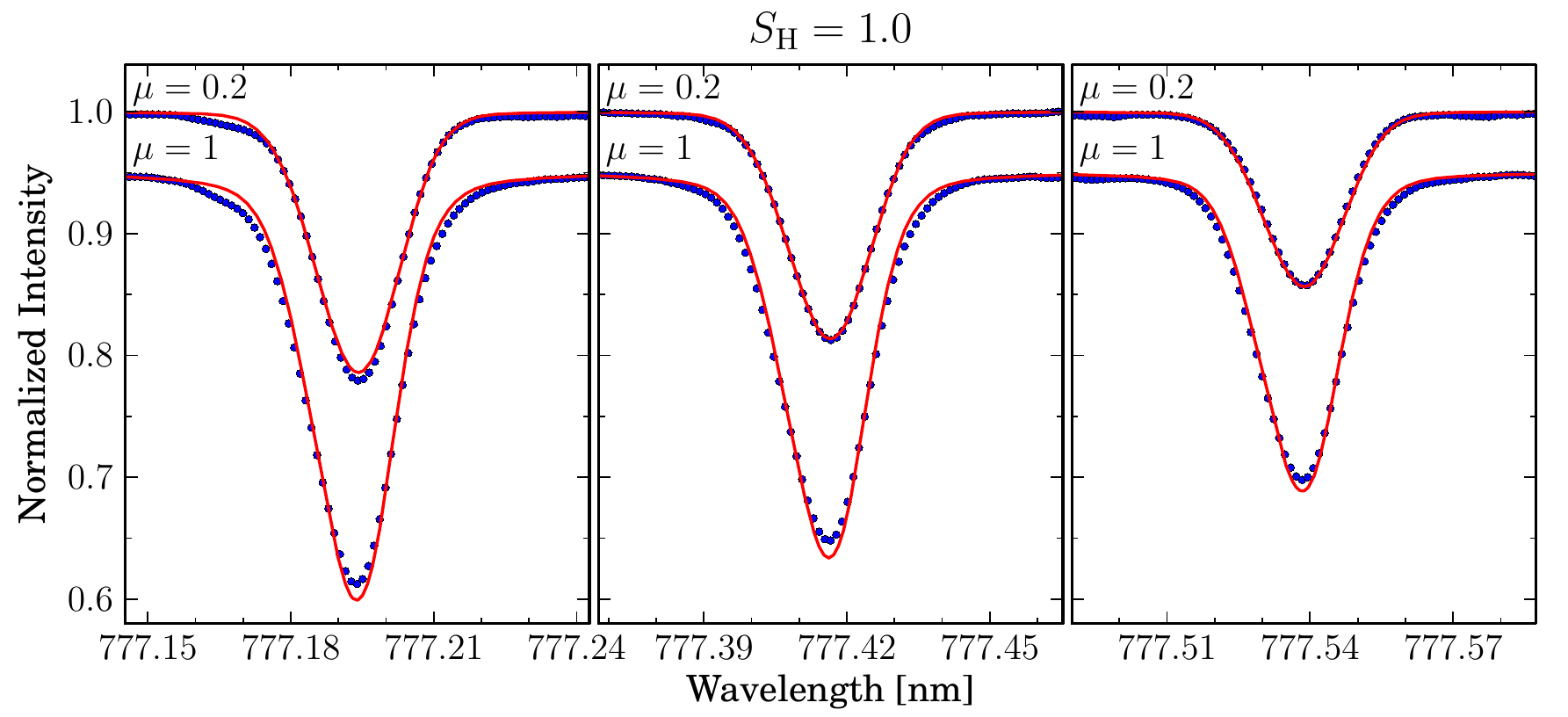}   \\
  \includegraphics[width=0.6\textwidth]{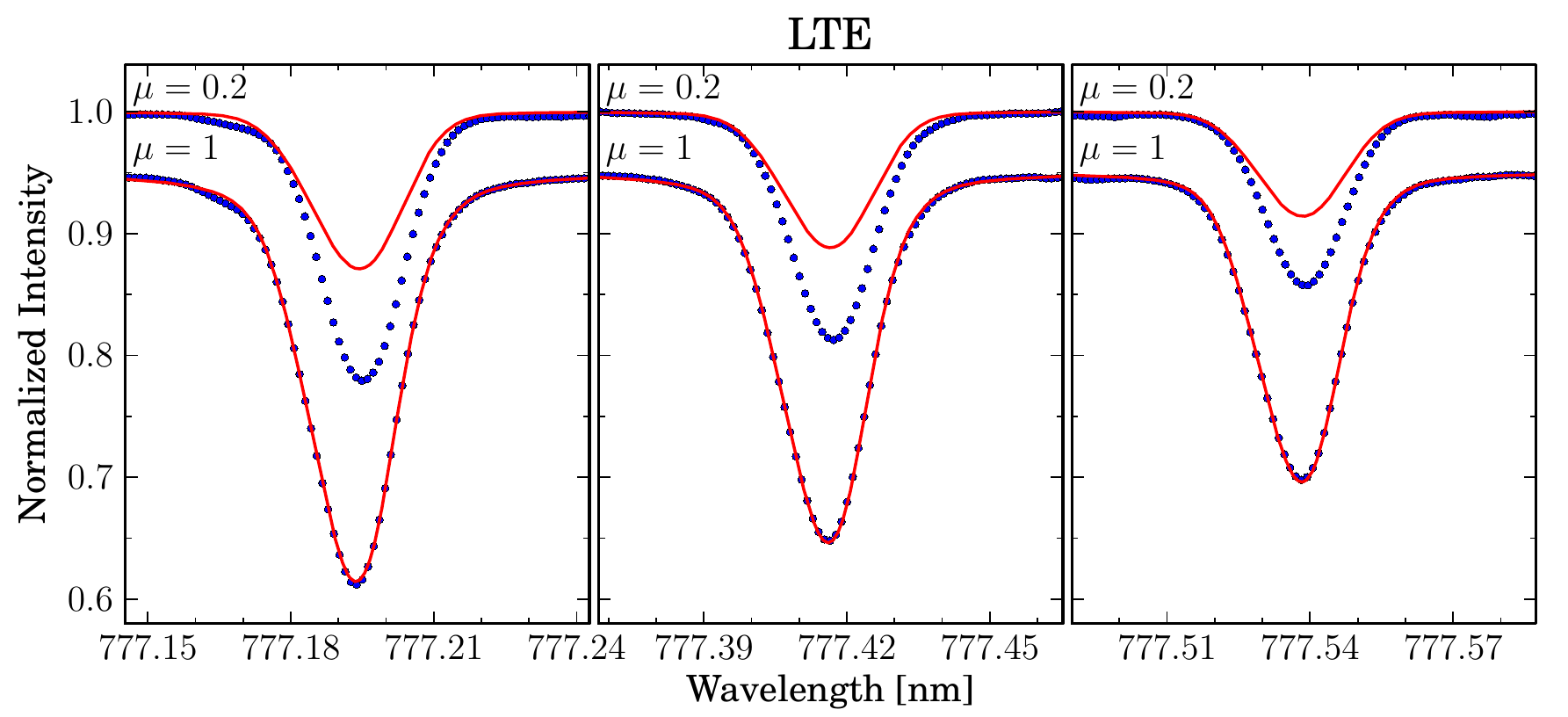}

\caption{Our observations (dots) and 3D synthetic line profiles (solid line). Each panel shows the 777~nm triplet lines at disk-centre and $\mu=0.2$, for LTE and three different $S_{\mathrm{H}}$ values. The line profiles at disk-centre were fitted to the observations, while at $\mu=0.2$ the abundance was set fixed to the disk-centre abundance, while the waveshift was adjusted. For clarity, disk-centre line profiles have been subtracted 0.05.}
  \label{fig:777_profs}
\end{figure*}

In Fig.~\ref{fig:777_profs} the 3D model predicted line profiles for disk-centre (adjusted in abundance) and the limb (using the disk-centre abundance), are shown for LTE and \mbox{$S_\mathrm{H} =$ 0.01, 0.3, and 1}. It can be seen that LTE performs very poorly at the limb (too weak), that $S_\mathrm{H}=0.01$ does not fare much better (too strong) and that $S_\mathrm{H}=1$ shows a much better agreement.

 One can also see that at disk-centre the LTE profiles seem to fit the observations better. The disk-centre profiles for $S_\mathrm{H}=1$ are narrower and deeper than the observed. At the limb, in terms of shape only, $S_\mathrm{H}=1$ profiles have a much better fit. The $S_\mathrm{H}$ that best describes the centre-to-limb variation of the line strengths does not seem to be the best at describing the shapes of line profiles at disk-centre. This same effect is more obvious when looking at the granulation variation of the FWHM at disk-centre \citepalias{Pereira2009b}. It may very well be connected with the finding of \citetalias{Pereira2009b} that no single $S_\mathrm{H}$ can provide the best agreement with the observed equivalent widths for both granular and intergranular regions. The reason for these discrepancies is yet not clear.

\begin{figure}
\centering
  \includegraphics[width=0.4\textwidth]{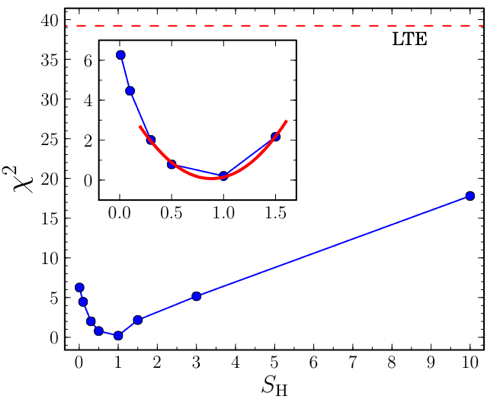}

\caption{Reduced $\chi^2$ as a function of $S_\mathrm{H}$ from the minimization of the equivalent widths (see text), for the 3D model. The LTE value is indicated (dashed line). The minimum at $S_\mathrm{H}=1$ indicates the best agreement with the observations. \emph{Inset:} same plot in more detail for $S_{\mathrm{H}}<2$, showing the parabolic fit to the lower $\chi^2$ values (thick solid line).}
  \label{fig:sh_min}
\end{figure}

To better quantify the agreement with the observations of different $S_\mathrm{H}$, a $\chi^2$ minimization was made with the equivalent widths, simultaneously for the three lines and for the five $\mu$ values. For each value of $S_\mathrm{H}$ we varied the oxygen abundance and found the value that minimizes the squared difference between observed and predicted equivalent widths, weigthed by the observational error bars. The reduced $\chi^2$ is defined as \mbox{$1/N\cdot\sum\left(W_{\mathrm{obs}}-W_{\mathrm{model}}\right)^2/\sigma^2$}, where $N$ is the number of degrees of freedom (in this case, $N=15-1$). The $\chi^2$ values, as a function of $S_\mathrm{H}$ are shown in Fig.~\ref{fig:sh_min}. Comparing for all the $S_\mathrm{H}$ we identify the value that gives the best agreement, using a parabolic fit to the lower $\chi^2$ values.  %

For the 3D model we find the $\chi^2$ minimum to be at \mbox{$S_\mathrm{H}\approx0.85$}. We use this value to derive the oxygen abundance from these lines. 
Averaging over the three lines we obtain an oxygen abundance of $\log\epsilon_{\rm O}=8.68$ for the 3D model. Repeating the procedure for the 1D Holweger--M\"uller model we also obtain $S_\mathrm{H}\approx0.85$, deriving an abundance of 8.66. Although with the 1D \textsc{marcs} model no $S_\mathrm{H}$ reproduces the observations, for comparison we derive an abundance of 8.61 using $S_\mathrm{H}=0.85$.

\subsubsection{Comparison with previous work}

At first glance our findings are consistent with the $S_\mathrm{H}=1$ estimated by \citetalias{CAP2004}. But these two findings cannot be directly compared, as we use a different 3D model. %
To compare our methods and observations with \citetalias{CAP2004}, we repeated our analysis for the same 3D model they use. The best fitting value is then $S_\mathrm{H}\approx 0.3$, which gives a similar agreement with the observations as $S_\mathrm{H}=1$ with the new 3D model. Between $S_\mathrm{H}=0$ and $S_\mathrm{H}=1$ (the two values tested by \citetalias{CAP2004})  we find that the latter gives a better agreement with the observations, consistent with the findings of \citetalias{CAP2004}. But the best fit of $S_\mathrm{H}\approx 0.3$ indicates a model dependence in this derivation, also seen for the 1D models. Interestingly, although the best fitting $S_\mathrm{H}$ varies for the two 3D models, the abundances derived from profile fits with the best choice $S_\mathrm{H}$ are almost unchanged between the two models ($\la 0.01\,\rm{dex}$).

\begin{table} 
\caption{Derived oxygen abundances ($\log\epsilon_\mathrm{O}$) from fitting disk-centre line profiles from the 3D model, using 3D NLTE radiative transfer for different values of $S_\mathrm{H}$ both for our SST observations and the FTS atlas of \citet{BraultNeckelFTS}.} 
\label{table:abund77} 
\centering 
\begin{tabular}{c l c c} 
\hline\hline
O\,\textsc{i} line [nm] & $S_\mathrm{H}$ & SST/TRIPPEL & FTS atlas \\ 
\hline 
777.19 & 0.01 & 8.50 & 8.52 \\ 
       & 0.1  & 8.53 & 8.55 \\ 
       & 0.3  & 8.58 & 8.60 \\
       & 0.5  & 8.61 & 8.63 \\ 
       & 1    & 8.66 & 8.68 \\ 
       & 1.5  & 8.73 & 8.75 \\ 
       & 3    & 8.76 & 8.77 \\
       & 10   & 8.82 & 8.84 \\
       & LTE  & 8.87 & 8.88 \\ 
\hline 
777.41 & 0.01 & 8.54 & 8.55 \\ 
       & 0.1  & 8.56 & 8.57 \\ 
       & 0.3  & 8.60 & 8.61 \\ 
       & 0.5  & 8.63 & 8.64 \\
       & 1    & 8.69 & 8.69 \\ 
       & 1.5  & 8.74 & 8.75 \\ 
       & 3    & 8.76 & 8.77 \\
       & 10   & 8.82 & 8.83 \\
       & LTE  & 8.85 & 8.86 \\ 
\hline
777.53 & 0.01 & 8.58 & 8.58 \\ 
       & 0.1  & 8.60 & 8.60 \\ 
       & 0.3  & 8.63 & 8.64 \\
       & 0.5  & 8.65 & 8.66 \\
       & 1    & 8.69 & 8.70 \\ 
       & 1.5  & 8.74 & 8.75 \\ 
       & 3    & 8.75 & 8.76 \\
       & 10   & 8.80 & 8.81 \\
       & LTE  & 8.83 & 8.83 \\ 
\hline\hline
\end{tabular} 
\end{table}

\subsubsection{Effect of blends and electron collisions on the inferred $S_\mathrm{H}$ and abundances}

There are a few weak CN and C$_\mathrm{2}$ lines in the 777~nm region. The VALD database also lists some weak atomic lines. Most of these lines are very weak and their effect on the O\,\textsc{i} lines is negligible. At disk-centre the known blends\footnote{There are a few unknown blends visible in the disk-centre spectrum. Most notably, the feature at the `knee' of the wing of the 777.19~nm line. Their effects in the fit were minimized by limiting the profile fitting region to parts not visibily affected by blends.} of the O\,\textsc{i} lines are very faint. But as molecular lines get stronger at the higher layers probed by the limb spectra, it is important to determine if they have an effect on the inferred $S_\mathrm{H}$ (the effect on disk-centre abundance obtained by profile fitting is negligible). To investigate for such effects we have computed the spectra of O\,\textsc{i} 777~nm over a larger spectral region and including additional molecular and atomic lines, with data from J. Sauval (2008, private communication). Our testing showed that the blends have very little influence in the derived $S_\mathrm{H}$ value. Because the atomic and molecular data for these weak blending lines is not known accurately, they have not been included in our main analysis.

We also investigate the effect of computing the O\,\textsc{i} lines simultaneously: their wings are very extended and overlap, especially in LTE and for 777.41 and 777.53~nm. For LTE at disk-centre, computing the three O\,\textsc{i} lines simultaneously caused a reduction of 0.002 dex in the fitted oxygen abundance of 777.41~nm line. At the limb there is virtually no difference between computing the lines separately or together, because the lines are much weaker. For the same line strength the LTE profiles have the most extended wings when compared with NLTE  (the extent of the wings gets smaller with smaller $S_\mathrm{H}$) and thus it is for the LTE profile that the issue of overlapping wings gets more important. 
Our single line approximation is thus fully justified.

In this work we make use of revised electron collision data \citep{Barklem2007}. Although this data is believed to be the most accurate currently available, it is important to check to what point our empirically estimated $S_\mathrm{H}$ depends on the adopted electron collision rates. To assess this effect, we have performed several tests by artificially changing the electron collisions. Due to time constraints these tests have been performed in 1D models, although they were confirmed with a few 3D runs. An interesting find is that if the electron collision rates are artificially decreased, the best fitting $S_\mathrm{H}$ value is still the same (even if we neglect electron collisions altogether). Obviously the fitted abundance values will differ (especially for  $S_\mathrm{H}<1$), but the best fitting value remains the same. This result indicates that H\,\textsc{i} collisions are more important than electron collisions in the centre-to-limb variation of the lines, because the electron pressure decreases more quickly with height. On the other hand, if the electron collision rates are increased by one order of magnitude it will change the best fitting $S_\mathrm{H}$ value. This makes sense, as an increase on the collisions from electrons will shift the result more towards LTE, regardless of the $S_\mathrm{H}$ used. %

\subsection{[O\,I] 630.03~nm and Ni\,I blend}
\subsubsection{Context}

Although the forbidden O\,\textsc{i} line at 630.03~nm is the strongest of the [O\,\textsc{i}] lines it is still very weak in the Sun. Unlike the permitted lines, LTE is a very good approximation in the formation of the [O\,\textsc{i}] line \citep[see][]{Asplund2004}. Accordingly, we only use the synthetic LTE profiles in the following analysis. Being a weak line it is highly susceptible to even faint blends. In particular, it is blended with a Ni\,\textsc{i} feature at 630.0335~nm \citep{Lambert1978,CAP2001,Johansson2003}. For the Sun, the contribution of the Ni\,\textsc{i} component to the blend is about 30\%, depending on the Ni abundance used. With such a significant blend it is of paramount importance to properly account for the Ni contribution. This work is similar to the work of \citet[hereafter ALA01]{CAP2001}, that included an analysis of [O\,\textsc{i}] 630.03~nm with a 3D model (albeit an earlier version), but for the flux profile instead of intensity at different $\mu$ values; also \citet{Caffau2008} and \citet{Ayres2008} have performed similar studies using a different 3D model.

\subsubsection{Ni\,\textsc{i} blend}

Properly accounting for the Ni\,\textsc{i} blend means one needs to know its properties with high precision. \citet{Johansson2003} have measured the $\log gf$ for this Ni\,\textsc{i} transition to be $-2.11$. Because of the presence of several Ni isotopes and their isotopic splitting, the Ni contribution at 630.03~nm is given as a series of blends. In our treatment we included the five most abundant Ni isotopes scaling the $\log gf$ by their isotopic abundance (see Table~\ref{table:ni_iso}). We have used a fixed Ni abundance of $\log\epsilon_{\mathrm{Ni}}=6.22$, as inferred from other Ni lines using the present 3D model \citep{Asplund2009}. %

\begin{table} 
\caption{Ni\,\textsc{i} isotopes used and scaled $\log gf$ values. Wavelength of main component and isotopic splitting from \citet{Johansson2003}, isotopic fractions from \citet{Rosman1998}.} 
\label{table:ni_iso} 
\centering 
\begin{tabular}{c r c c} 
\hline 
Ni isotope & \% & Wavelength [nm] & $\log gf$ \\
\hline\\
   ${}^{58}$Ni    & 68.27  & 630.0335  & -2.2757 \\ 
   ${}^{60}$Ni    & 26.10  & 630.0355  & -2.6933 \\ 
   ${}^{61}$Ni    &  1.13  & 630.0365  & -4.0569 \\ 
   ${}^{62}$Ni    &  3.59  & 630.0375  & -3.5549 \\ 
   ${}^{64}$Ni    &  0.91  & 630.0395  & -4.1509 \\ 
\hline
\end{tabular} 
\end{table} 

An important difference between the present work and \citetalias{CAP2001} was the treatment of the Ni\,\textsc{i} blend. We used a fixed Ni abundance of 6.22 and adopt the measured $\log gf$ of $-2.11$, whereas at the time of \citetalias{CAP2001} there was no experimental $\log gf$ and Ni was treated as a free parameter. Their best fitting $\log(gf\epsilon_{\mathrm{Ni}})=3.94$ is equivalent to a Ni abundance of 6.05 with the present value of $\log gf$, smaller than the Ni abundance we use. In \citetalias{CAP2001} the blend of the O\,\textsc{i} and Ni\,\textsc{i} was done by co-adding the individual flux profiles for the different species. Because these are very weak lines, that was a reasonable approximation. However, the approximation of co-adding the profiles starts to break down as one looks at profiles of smaller $\mu$ values, where the difference between co-adding and a proper treatment of the blend (summing the opacities for each component) can amount to a few percent of the equivalent width. In this work we opted to treat the blend properly, instead of co-adding the profiles.

\subsubsection{Wavelength calibration}

Another important point in the treatment of this line is the wavelength calibration. As \citet{Ayres2008} noted, the balance between the components of the blend is sensitive to velocity errors of a few hundred {m\,s$^{-1}$}, or a few {m\AA}. If the wavelength calibration translates the line to be blueshifted from its `true' location the fitting will favour a higher oxygen abundance. There is a degeneracy between the wavelength shift and the Ni abundance, which makes it difficult to determine the best-fitting values when both are allowed to vary.

Unfortunately the absolute wavelength calibration of neither our observations nor the FTS atlas is precise enough for the analysis of this blend. Hence we adopt a correction to the laboratory wavelength similar to that of \citet{Ayres2008}: using strong Fe\,\textsc{i} lines in the neighbourhood to find the wavelength difference between the 3D model and the observations. By measuring the wavelength difference between the synthetic profiles and the observations for a set of reference Fe\,\textsc{i} lines, one can effectively put the profiles in the same `laboratory' wavelength scale. This approach has the advantange of removing the systematics associated with the observations (\emph{e.g.} solar rotation) and any shortcomings in the 3D model (\emph{e.g.} overshooting of convective blueshifts). We chose the three Fe\,\textsc{i} lines at 629.77, 630.15 and \mbox{630.24~nm} -- as they were available in the limited wavelength window of our observations -- to calibrate the wavelength scale. To measure the wavelength difference we fit the theoretical 1D and 3D line profiles of these Fe\,\textsc{i} lines and allow a wavelength shift of the observations as a free parameter. 
The mean wavelength shift from the three lines was extracted for line profiles at all the solar disk positions of our observations. The standard deviation between the shifts from the three Fe\,\textsc{i} lines were 19, 27, 33, 51, 140 m\,s$^{-1}$ for respectively $\mu = 1, 0.8, 0.6, 0.4, 0.2$. Except the $\mu=0.2$ measurement, the numbers are an encouraging sign of how precise the wavelength calibration is. For the observations at $\mu=0.2$, due to the scatter between Fe\,\textsc{i} line indicators we adopted the wavelength shift as given by the Fe\,\textsc{i} 629.77~nm line only.

If the wavelength shift of the synthetic line profile was allowed to be a free parameter in the fit, then the best fitting shift to the FTS intensity atlas is very close ($<75\;\;\mathrm{m}\:\mathrm{s}^{-1}$) to the fixed shift obtained by the Fe\,\textsc{i} lines procedure. Its effect on the derived abundance is thus small ($< 0.01$ dex). Even if one accounts for some uncertainty around the wavelength shift, the effects for the oxygen abundance are small. A relatively large variarion of ($\pm 330\;\mathrm{m}\:\mathrm{s}^{-1}$) around the used waveshift for the FTS intensity atlas has an effect on the oxygen abundance of $\approx \mp 0.015$~dex, althought the agreement with the observed profile deteriorates significantly. These variations of the oxygen abundance are much smaller than the ones found by \citetalias{CAP2001} when trying different shifts. The difference is likely to be due to the fact that we kept the nickel abundance fixed, while \citetalias{CAP2001} allowed it to be a free parameter in the fit. With nickel fixed, the fitting procedure cannot compensate with more or less nickel, which means that the line shifting has a smaller effect on the oxygen abundance.

Besides the absolute wavelength calibration of the observations, there is also the issue of the uncertainties in the laboratory wavelengths for the oxygen and nickel transitions. The total uncertainty in the wavelength difference between the lines is approx. $\pm0.236$ pm, being $\pm 0.2$ pm for [O\,\textsc{i}] \citep{Eriksson1965}, and $\pm 0.125$ pm for Ni\,\textsc{i} \citep{Johansson2003}, weighted by the isotopes. We tested the line profile fitting for the lines using the two extreme cases of wavelength differences allowed by the errors and found that its effect on the derived oxygen abundance is negligible ($< 0.01$ dex). Once again, this value is smaller than the $0.05$~dex found by \citetalias{CAP2001}, because we do not have the nickel abundance as a free parameter.

\subsubsection{Continuum level and fitting range}

The local continuum of the [O\,\textsc{i}] + Ni\,\textsc{i} line wings is slightly depressed by the presence of nearby Fe\,\textsc{i} and Si\,\textsc{i} lines, and also about 70 weak CN lines between 630.00--630.06~nm, as noted by \citetalias{CAP2001}. To account for this effect, we multiplied each set of observations by a factor $C$. For the FTS intensity atlas $C=1.0054$ was used. It is important to note that the determination of this local continuum is a major source of the total uncertainty. If, for example, one just took the FTS intensity atlas's continuum and computed the equivalent width for this line, the equivalent width would be $\approx 30$\% higher. Care was taken in determining the local continuum for our observations, but one should keep in mind the uncertainty attached to it.

Another important but seldom discussed point in the profile fitting is the wavelength range where to fit the profile. \citetalias{CAP2001} have done the profile fitting in the range 630.015--630.04~nm, others are not so explicit. Ideally one would like to perform the profile fitting in a range as broad as possible to include the maximum amount of information about the line. The presence of blends, especially unknown features, forces one to narrow the fitting range to avoid the influence in the fit of features not included in the line synthesis. We have adopted the same wavelength range as \citetalias{CAP2001}, as it avoids several weak blends but includes most of the [O\,\textsc{i}] line.

\subsubsection{Comparison with observations}

\begin{figure*} 
\begin{center}
  \includegraphics[width=0.33\textwidth]{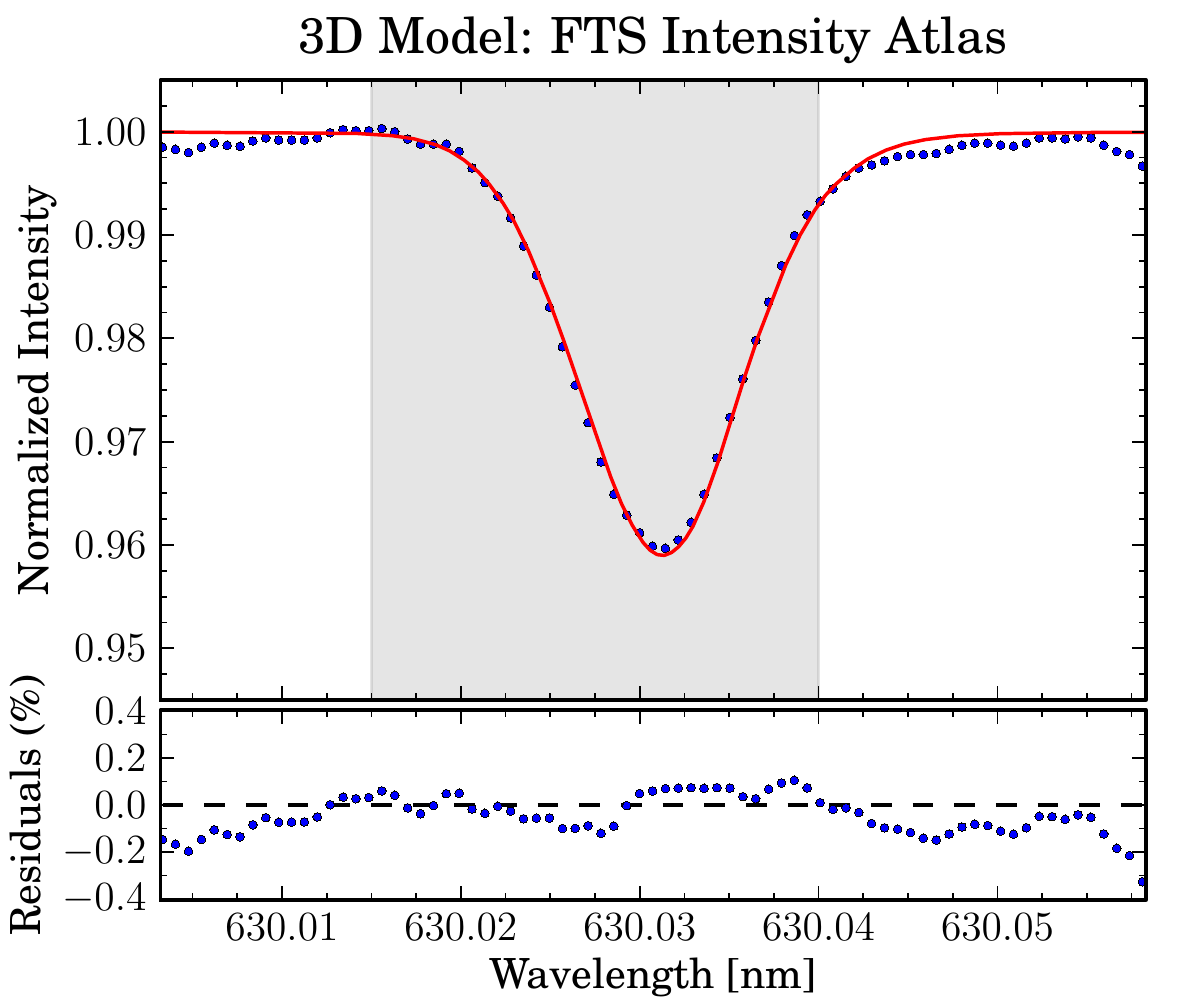}
  \includegraphics[width=0.33\textwidth]{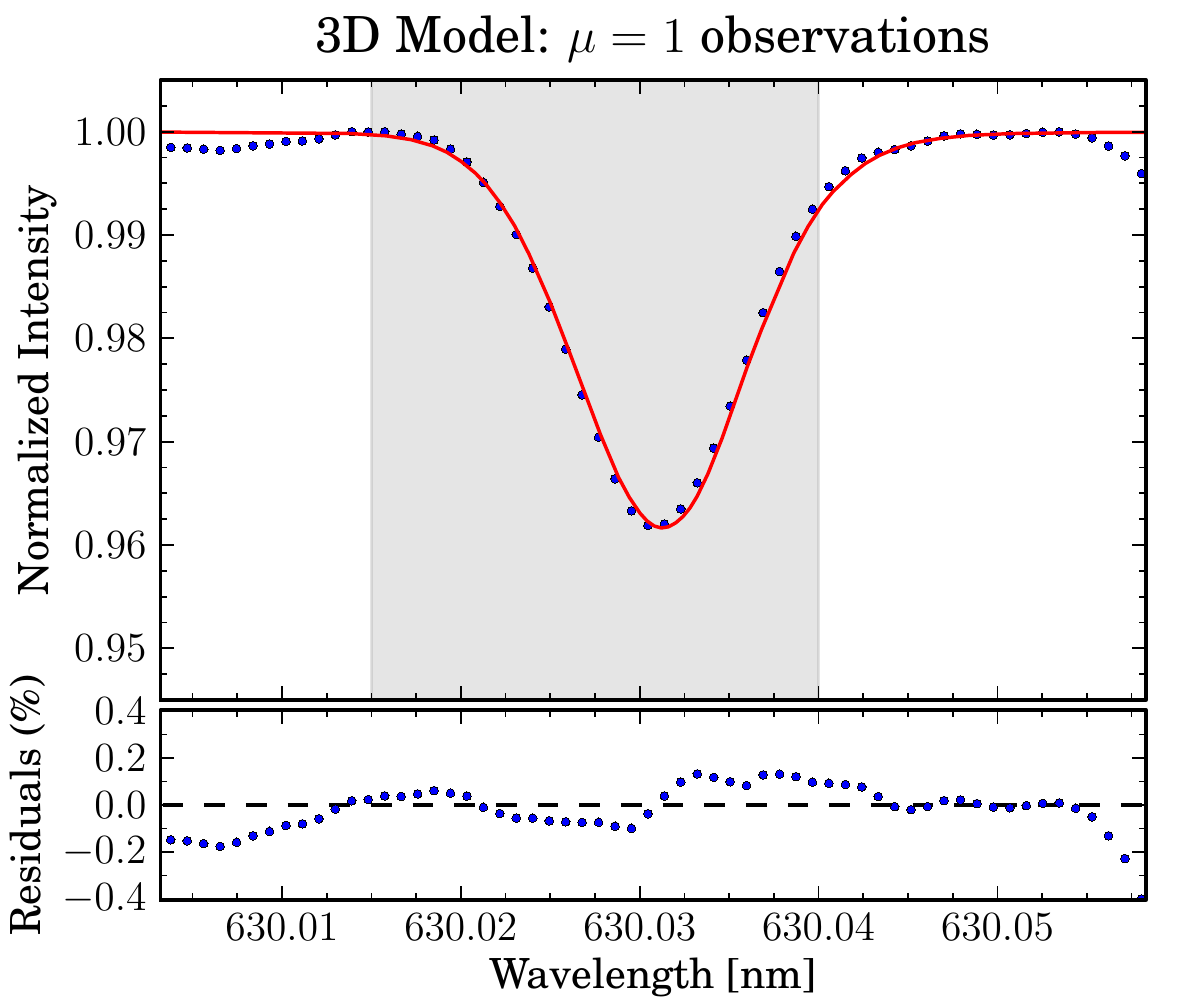}
  \includegraphics[width=0.33\textwidth]{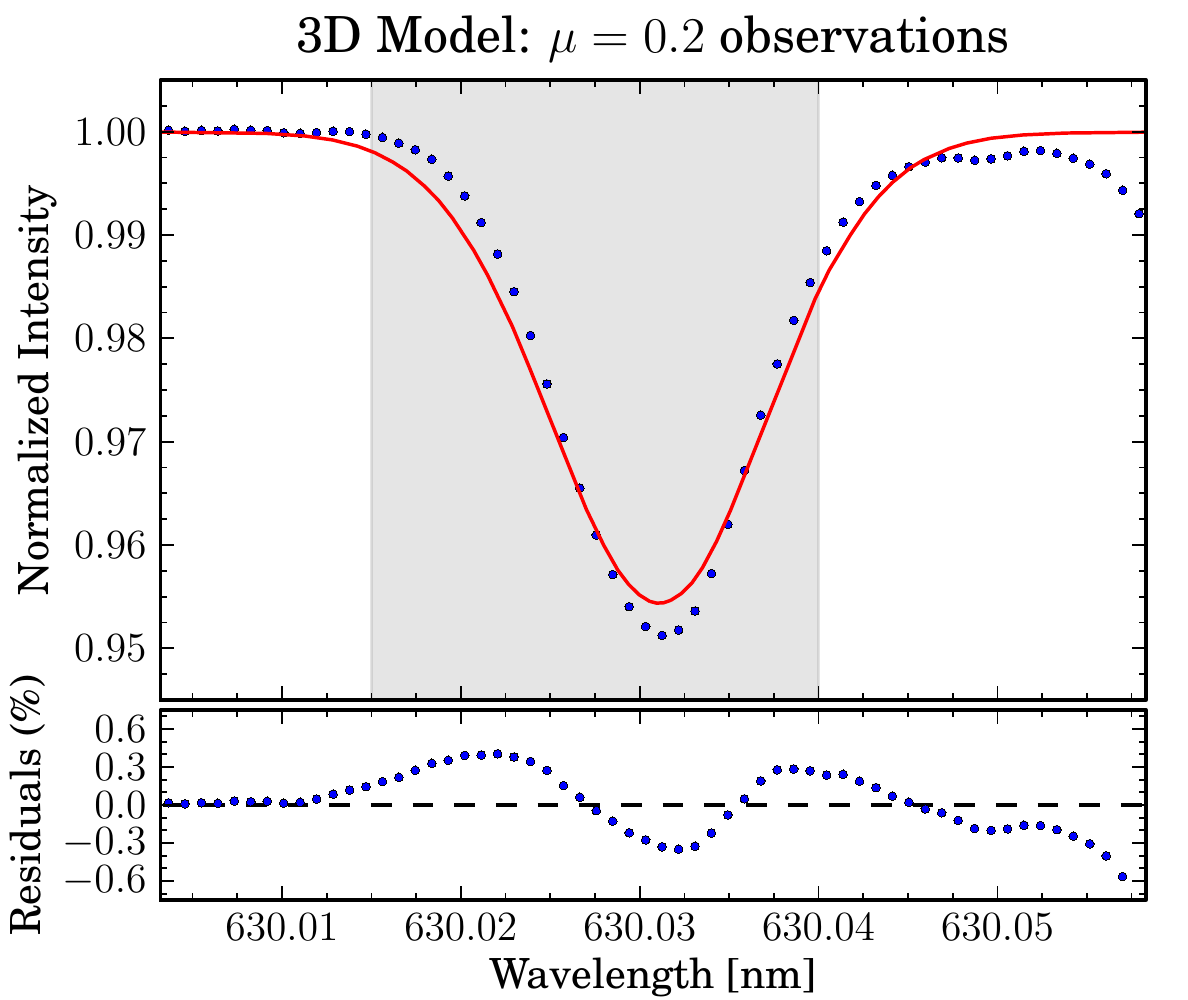}\\

  \includegraphics[width=0.33\textwidth]{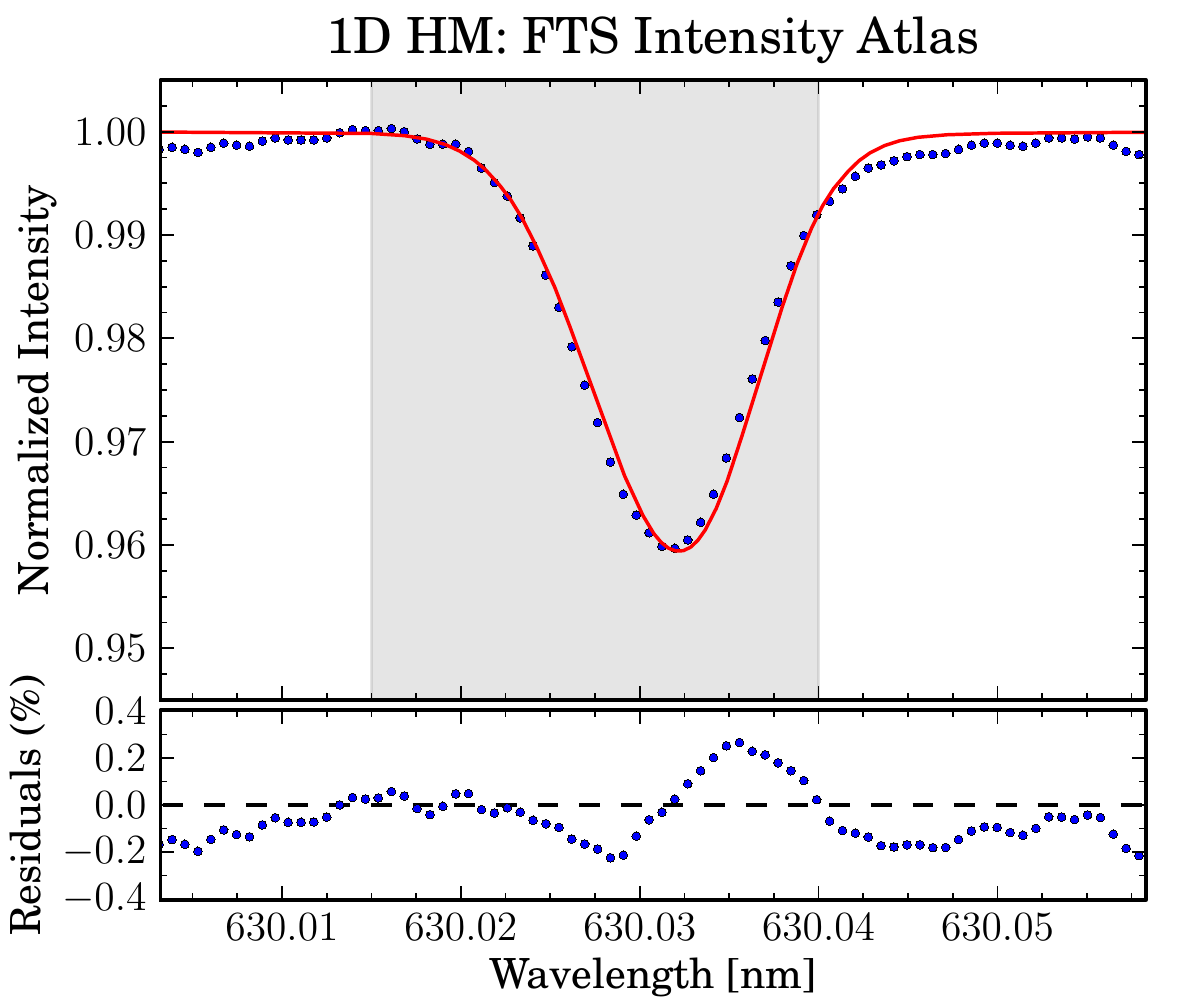}
  \includegraphics[width=0.33\textwidth]{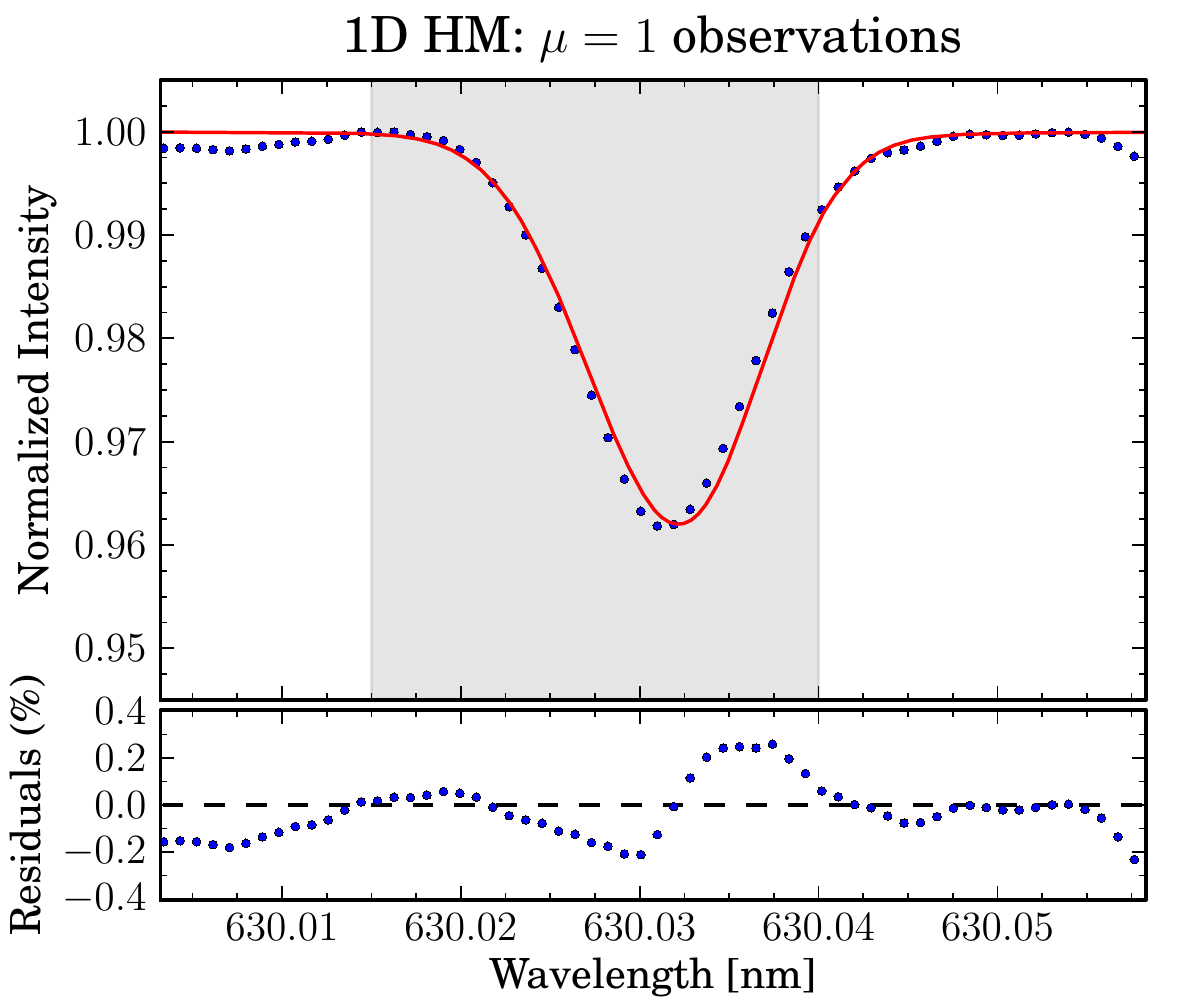}
  \includegraphics[width=0.33\textwidth]{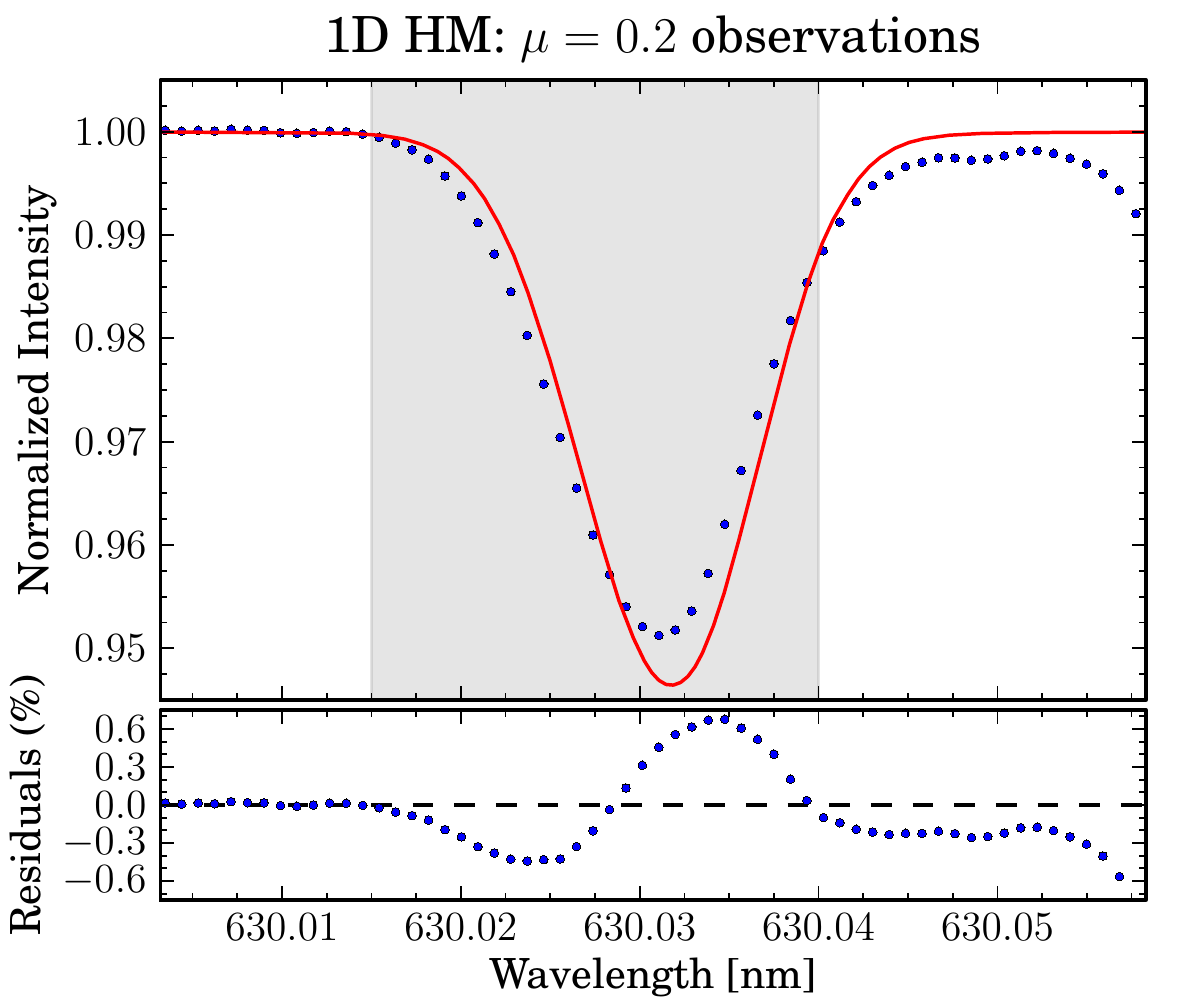}\\

  \includegraphics[width=0.33\textwidth]{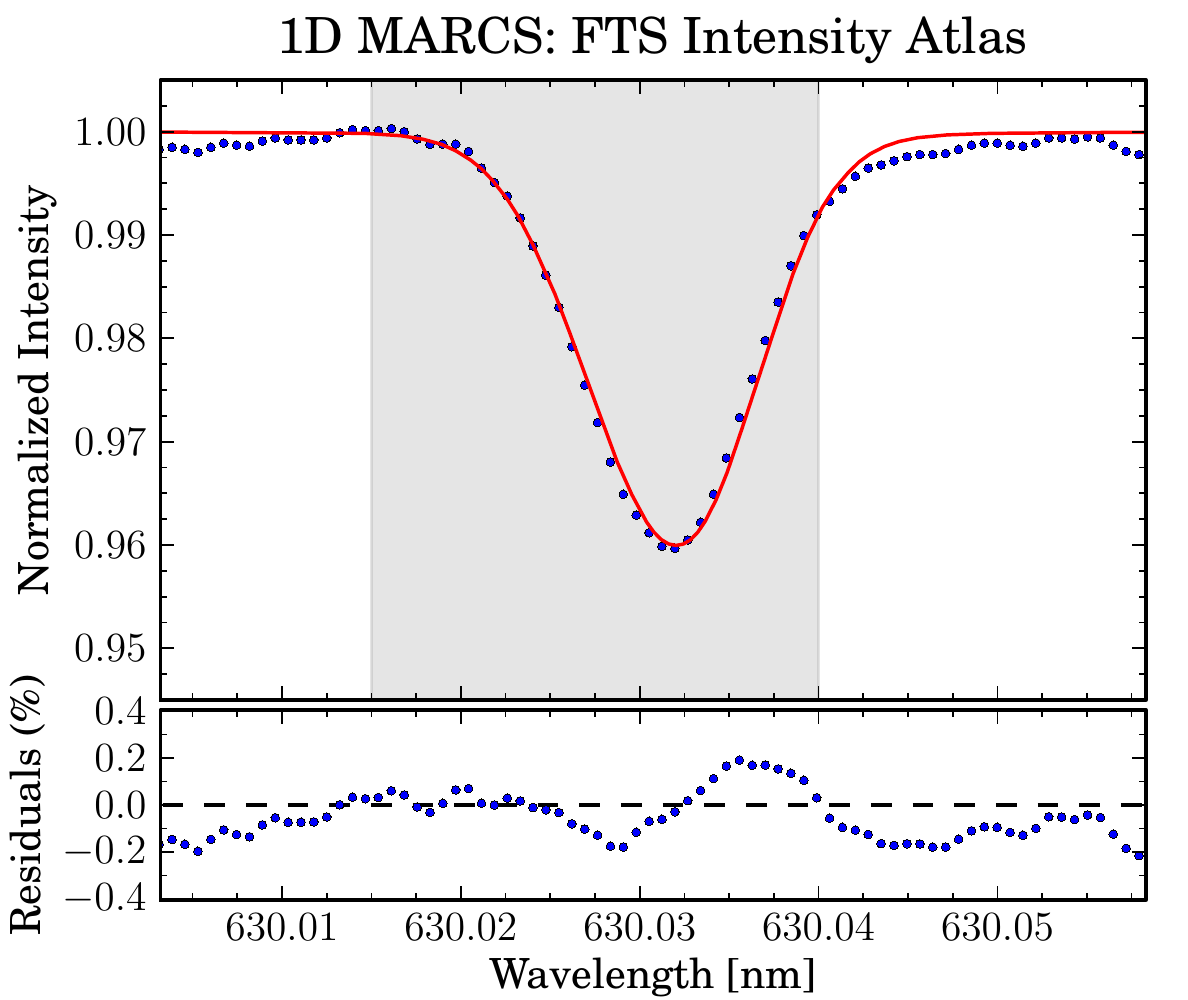}
  \includegraphics[width=0.33\textwidth]{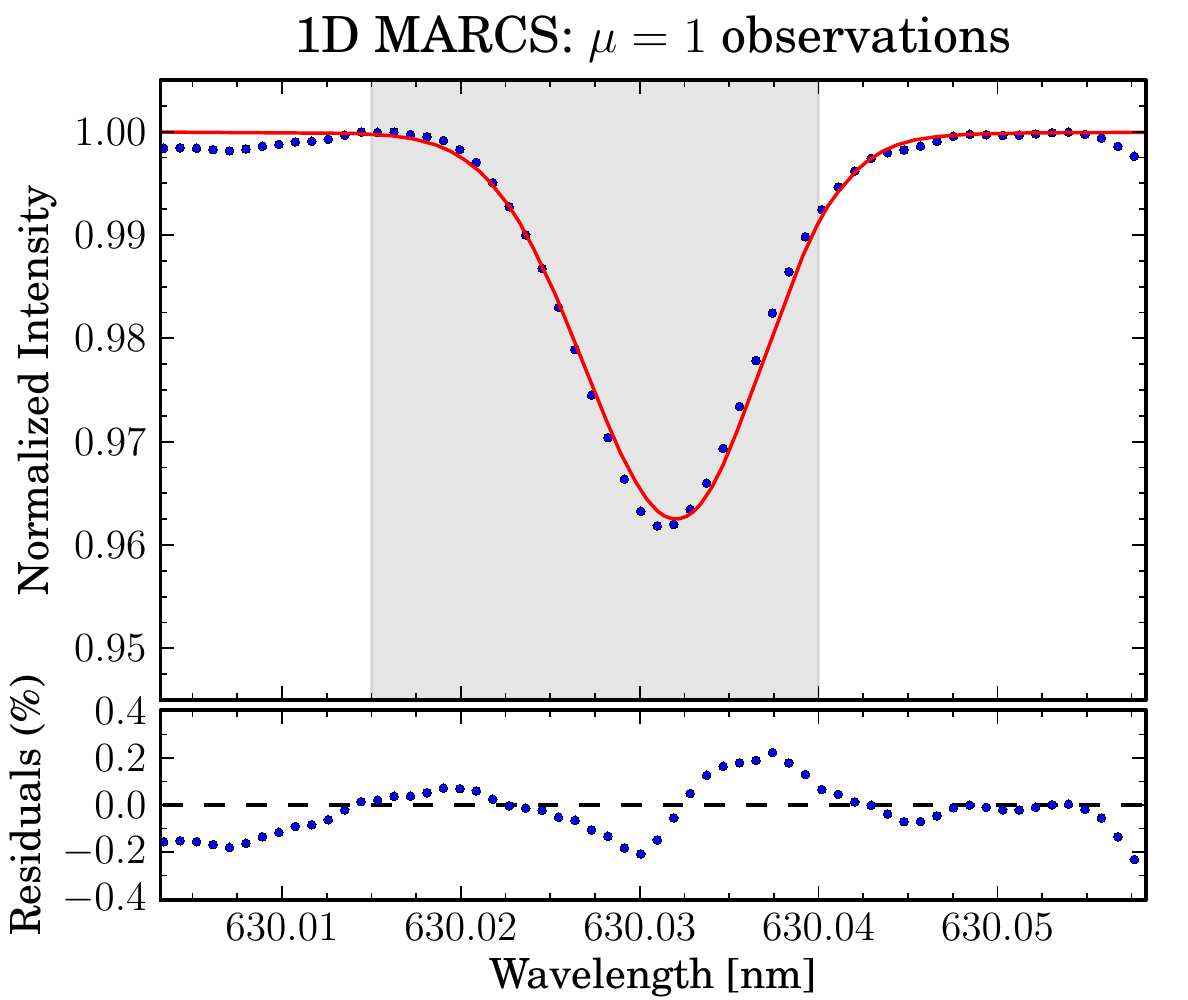}
  \includegraphics[width=0.33\textwidth]{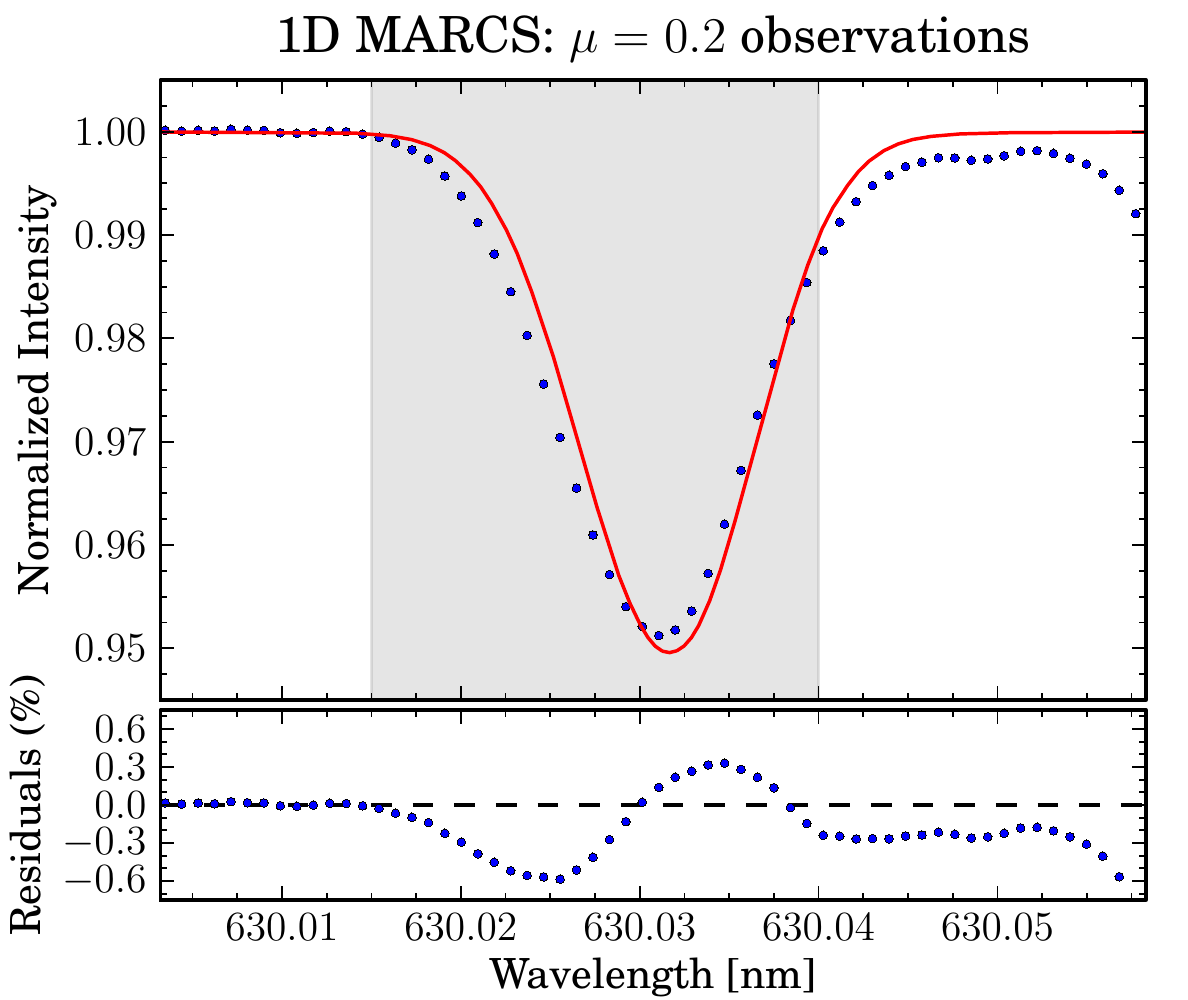}
\end{center}
\caption{Synthetic line profiles (solid lines) for the O\,\textsc{i}+Ni\,\textsc{i} 630.03~nm blend vs. observations (circles). Shaded regions indicate range over which the profiles were fitted. Left panels: fit for FTS Intensity atlas, done over the wavelength range of Allende-Prieto et al. (2001). Middle panels: fit for our observations, disk-centre. Right panels: our observations at $\mu=0.197$, and profiles from models using abundance fitted at disk-centre, adjusted for continuum. The wavelength shift was measured for each model and disk position separately from Fe\,\textsc{i} lines (see text). Ni abundance was fixed at 6.22, 6.16, 6.26 for the 3D model, 1D MARCS model and 1D HM model, respectively.}
  \label{fig:630_profs}
\end{figure*}

The line profiles for the [O\,\textsc{i}] + Ni\,\textsc{i} blend were computed for a fixed $\log\epsilon_{\mathrm{Ni}}$ and assuming LTE. In the fit for this line profile we only allowed one free parameter: the oxygen abundance. The waveshift for each position was obtained by the \mbox{Fe\,\textsc{i}} lines method, and the local continuum multiplication factor $C$ obtained empirically, as noted above.

\begin{figure} 
  \includegraphics[width=0.45\textwidth]{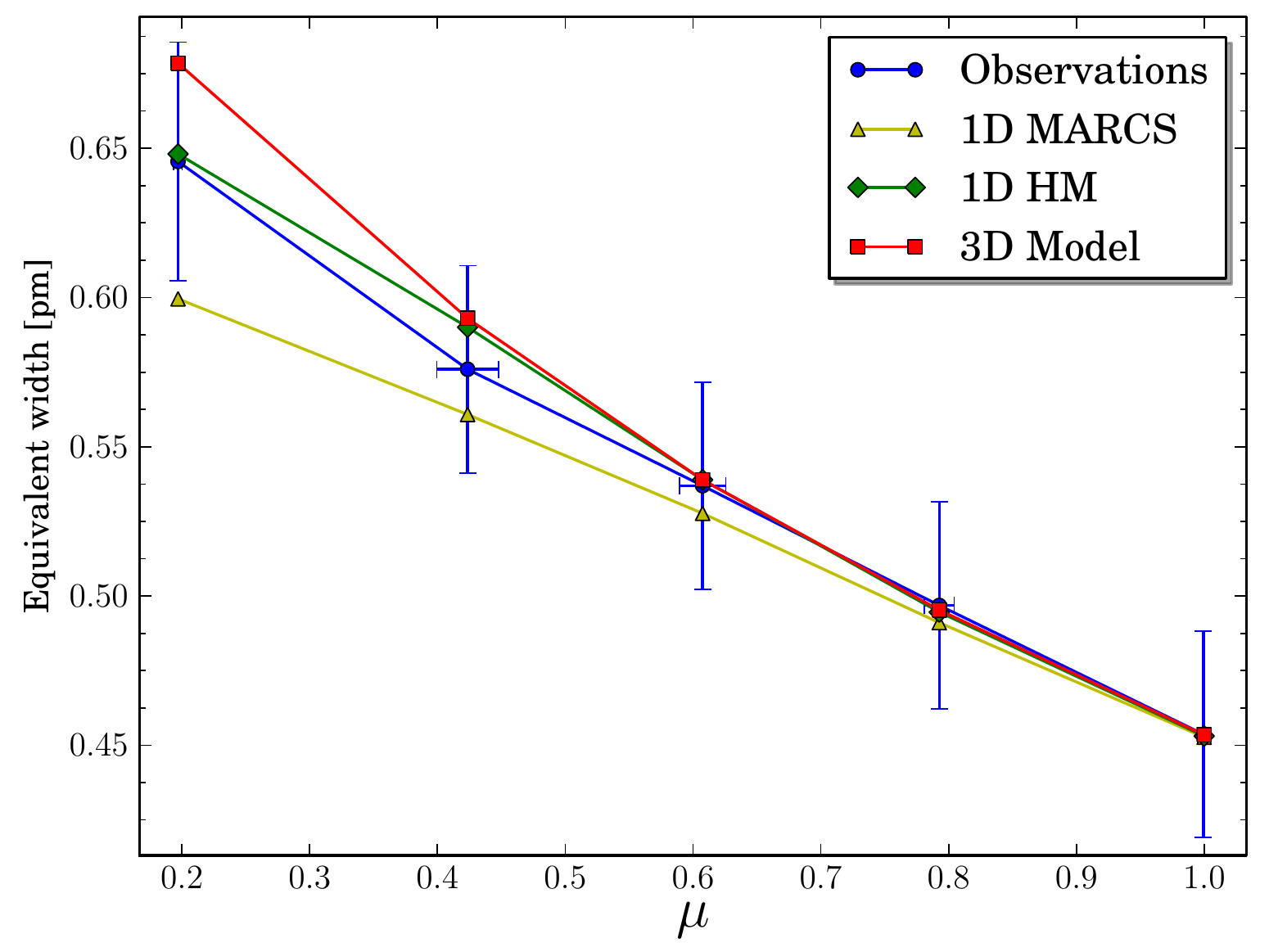}
\caption{Equivalent width vs. $\mu$ for the [O\,\textsc{i}]+Ni\,\textsc{i} 630.03~nm lines blend, for our observations and different models at different positions in the solar disk.}
  \label{fig:630_eqw}
\end{figure}

We present the results for the equivalent width analysis over the observed positions in the solar disk, the profile fitting of our observations at $\mu=1,\:0.2$ and the FTS intensity atlas for comparison with other studies. Some results from line fitting are in Fig.~\ref{fig:630_profs}, and the equivalent width results are in Fig.~\ref{fig:630_eqw}. Fitting the line profile to the FTS intensity atlas we obtain an oxygen abundance of $\log\epsilon_{\rm O}=8.66$, while for our observations at $\mu=1$ we obtain 8.64. In the variation of equivalent width with $\mu$ in Fig.~\ref{fig:630_eqw} one can see that the 3D model predicts a slightly stronger line as $\mu$ decreases. In terms of fitting the line profiles for different $\mu$ the variation in fitted abundance for the 3D model is never more than 0.02 dex in regards to the disk-centre abundance. In the rightmost column of Fig.~\ref{fig:630_profs} we plot the observations at $\mu=0.2$ with a synthetic profile using the oxygen abundances derived from the disk-centre fits (our observations), the continuum level adjusted independently. The 3D model at $\mu=0.2$ is slightly stronger than the observations, and the line profile is slightly wider and not as deep as the observations. Still, in terms of best-fitting abundance the difference is minimal.

For the 1D models we have applied the same wavelength correction with Fe\,\textsc{i} lines, repeating the procedure for line profiles synthesized with the 1D Holweger--M\"{u}ller and \textsc{marcs} models. The best fitting macroturbulence was also extracted from these Fe\,\textsc{i} lines, and then used for O\,\textsc{i}+Ni\,\textsc{i}. The Holweger--M\"{u}ller (HM) model does a good job at describing the line strengths, as evidenced by the equivalent widths in Fig.~\ref{fig:630_eqw}. However, a closer inspection at the line profiles shows a red wing that is too strong. This may possibly indicate an excess of nickel: for this model $\log\epsilon_{\mathrm{Ni}}=6.26$ was used, as derived with the same model using other Ni lines. Looking at Fig.~\ref{fig:630_profs}, in particular at the residuals, it is clear that the agreement is not as good as for the 3D model. This is expected, as the 1D models do not reproduce the convective motions in the photosphere as well as the 3D model, even using micro and macroturbulences. In particular for the HM at $\mu=0.2$ the macroturbulence derived from the Fe\,\textsc{i} lines is probably low, as the line profile is noticeably deeper than the observations, while the equivalent widths are very similar. For the \textsc{marcs} model the trend in equivalent width vs. $\mu$ is different: the model predicts slightly weaker line strengths than the observations. The fitted line profiles at disk-centre also indicate an excess of intensity in the red wing. The line profile at low $\mu$ with the disk-centre abundance shows a slightly weaker line than the observations. The oxygen abundances from profile fitting with the FTS disk-centre atlas for the HM and \textsc{marcs} models were, respectively, 8.69 and 8.66.

\subsubsection{Comparison with previous work}
An easy comparison to be made is with \citetalias{CAP2001}. Some things have been done differently in the present study. We compute a proper blend of the two transitions (adding line opacities), instead of computing separate line profiles for each line and co-adding them. We also perform a more precise wavelength calibration usinging Fe\,\textsc{i} lines. And finally, we employ a more recent 3D Model and a higher Ni abundance. Fitting the flux profiles against the FTS flux atlas and using the same fitting range in \citetalias{CAP2001}, we obtain an oxygen abundance 0.07 dex lower. The reason for this difference is mainly due to a higher Ni abundance used (6.22 vs. 6.05), which translates to a \mbox{$\approx -0.07$ dex} difference in the derived oxygen abundance. The proper blending of the line instead of co-adding profiles causes a difference of $-0.01$ dex in oxygen. The proper wavelength calibration using Fe\,\textsc{i} lines instead of only correcting for gravitational redshift causes a difference of $+0.01$ dex.

Fitting the line profiles with $\log\epsilon_{\rm Ni}=6.22$ instead of $6.05$ leads to a better agreement with the observed disk-centre profile ($\chi^2_{6.22} \approx 0.53 \chi^2_{6.05}$), but a worse agreement for the observed flux profile ($\chi^2_{6.22} \approx 1.51 \chi^2_{6.05}$).

\subsection{O\,I 615.81~nm}

\subsubsection{Context}

\begin{figure*} 
\begin{center}
  \includegraphics[width=0.33\textwidth]{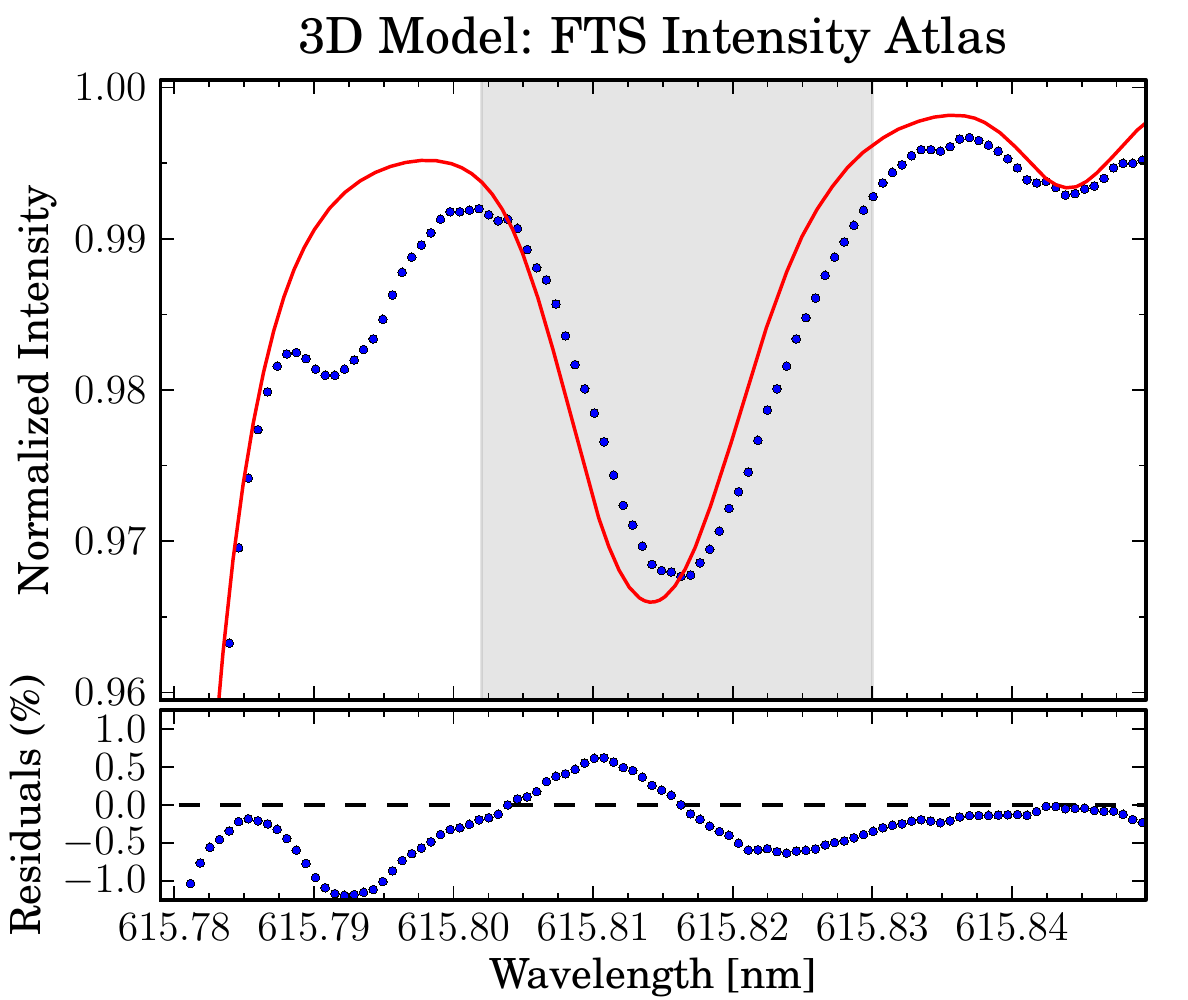}
  \includegraphics[width=0.33\textwidth]{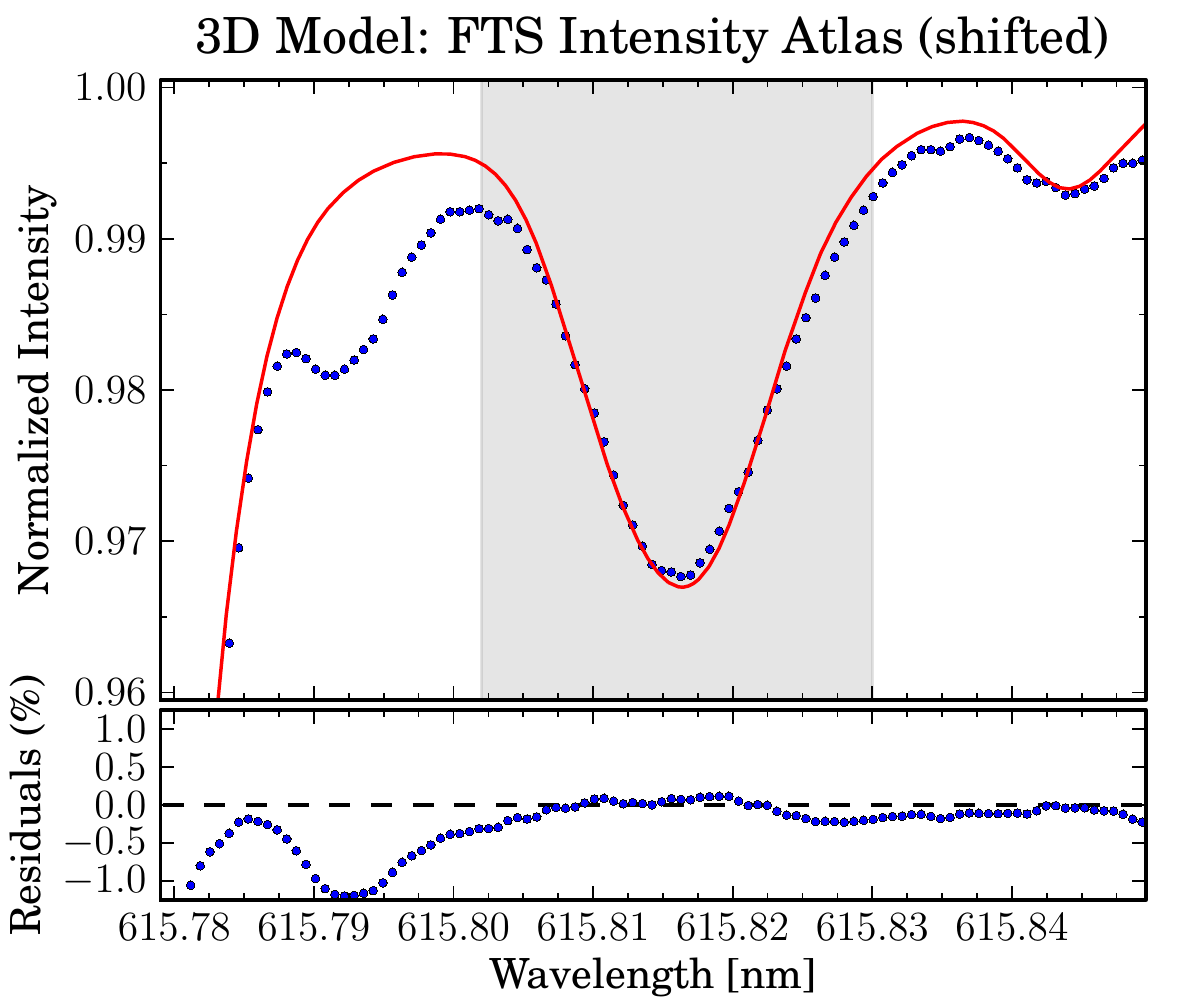} 
  \includegraphics[width=0.33\textwidth]{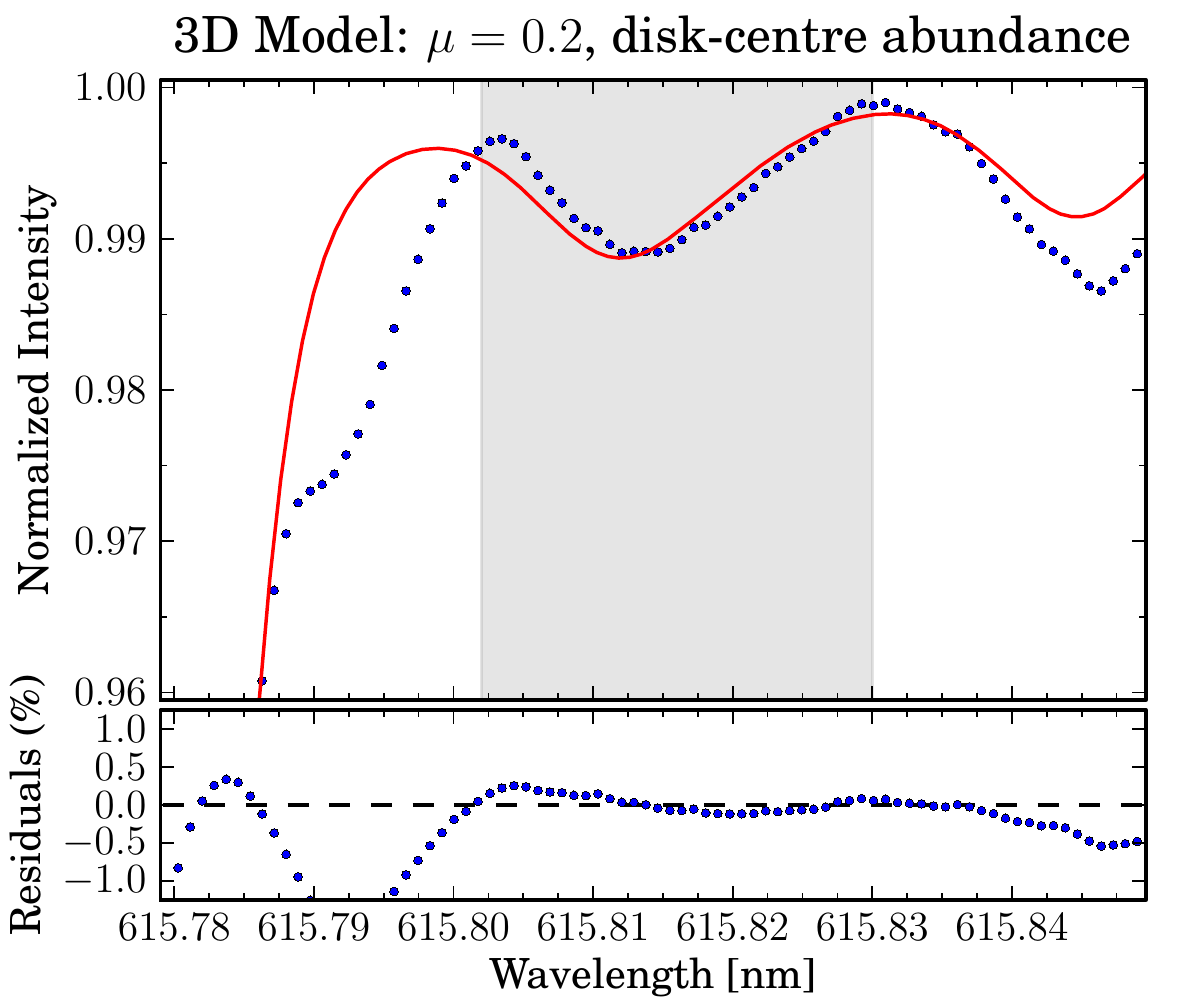} \\

  \includegraphics[width=0.33\textwidth]{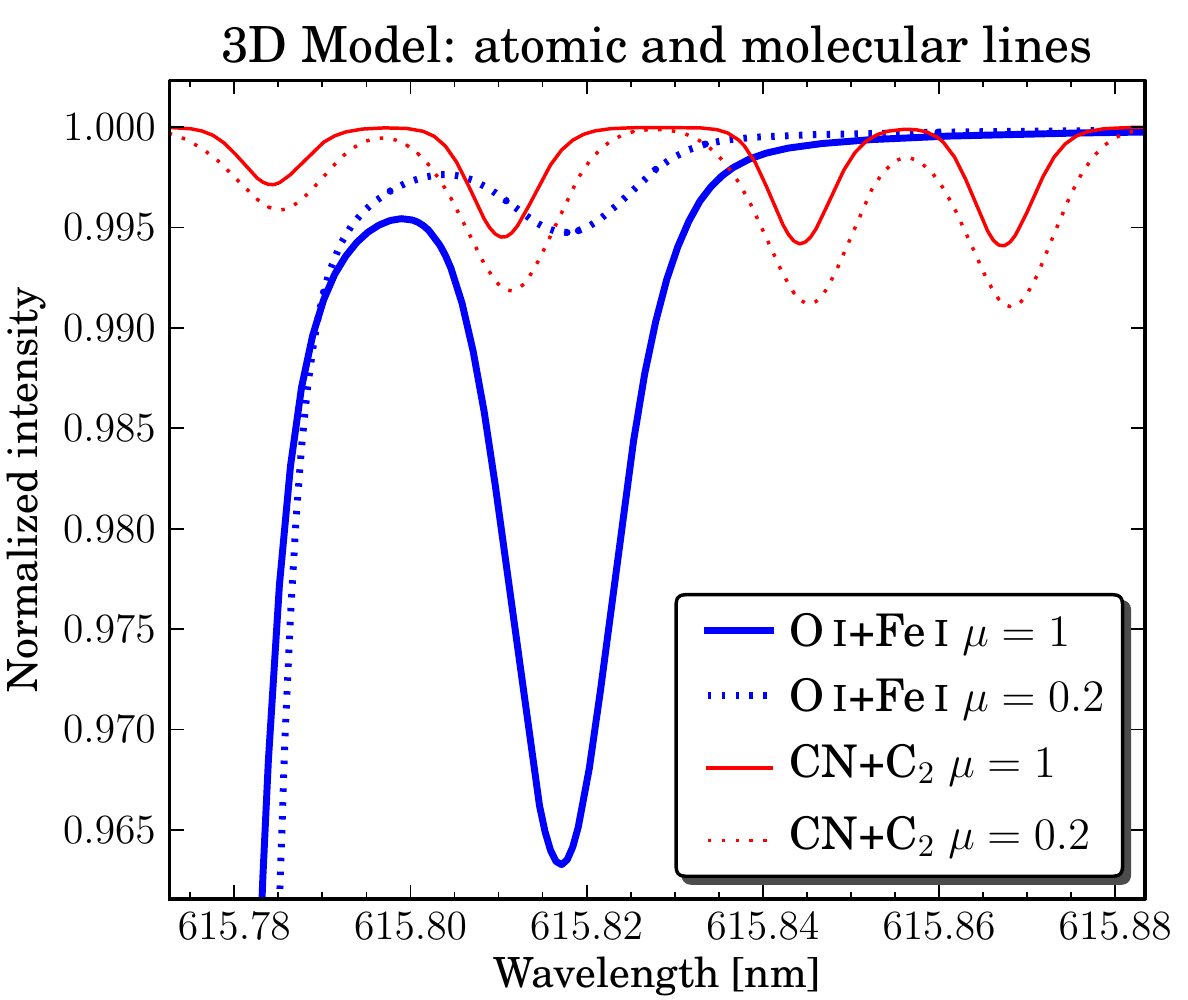} 
  \includegraphics[width=0.33\textwidth]{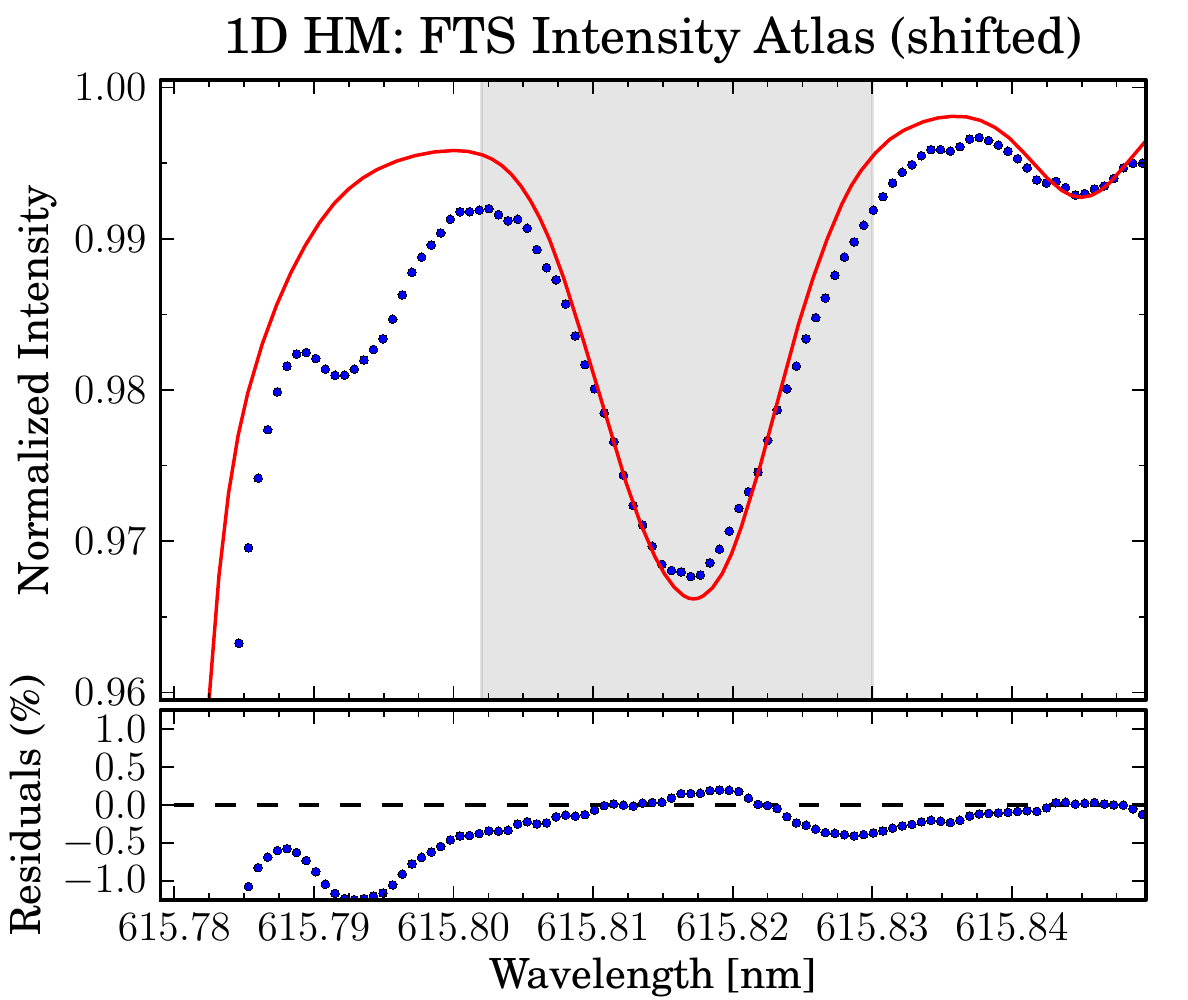} 
  \includegraphics[width=0.33\textwidth]{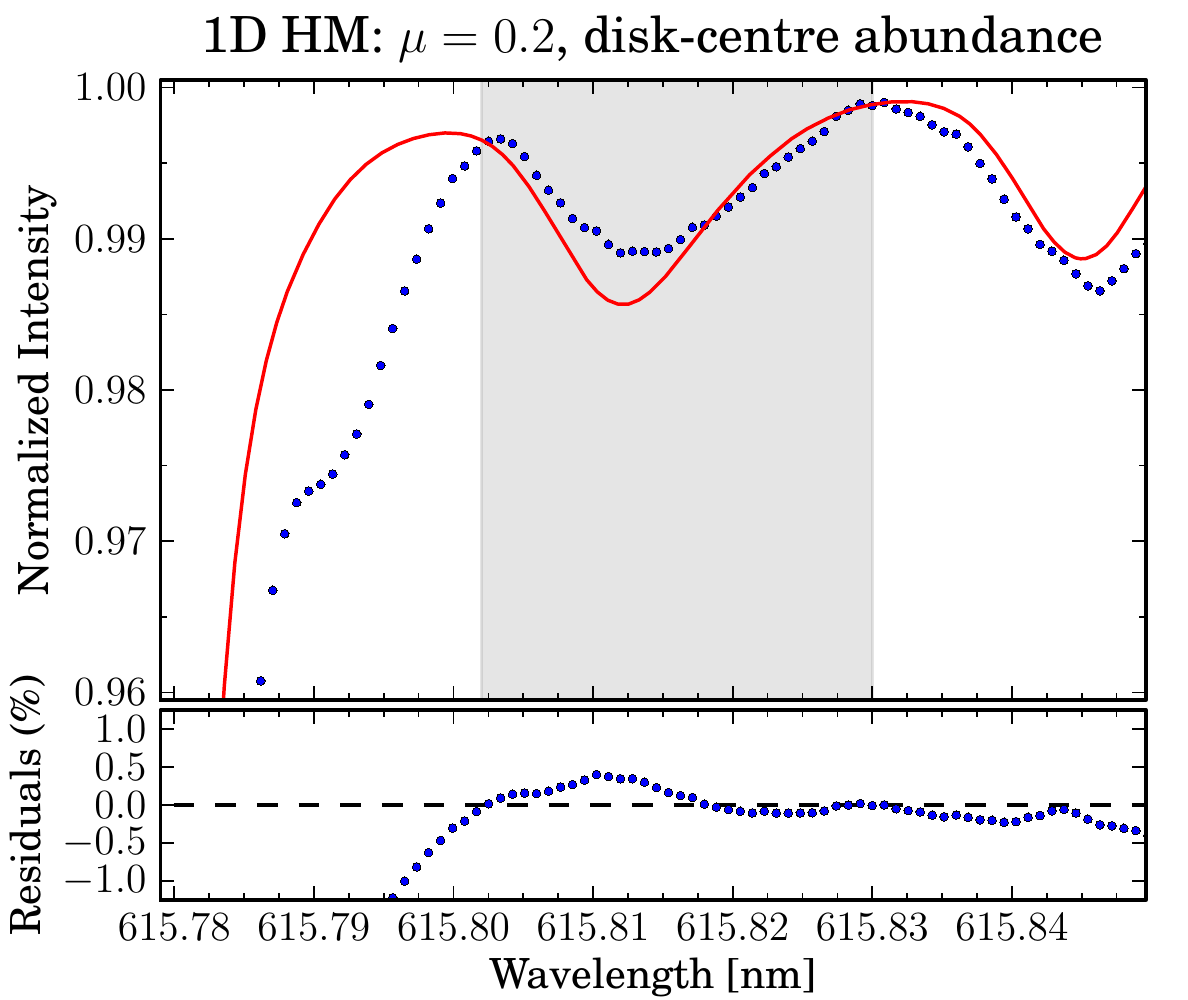} 
\end{center}
\caption{Synthetic LTE line profiles for the O\,\textsc{i} 615.81~nm lines (including blends from Fe\,\textsc{i}, one C$_\mathrm{2}$ and five CN lines). All panels except lower left show synthetic profiles (solid) vs. observations (circles). Top left panel: location of the O\,\textsc{i} line in respect to the observations, using waveshift from Fe\,\textsc{i} lines and laboratory wavelengths from the NIST database. Middle panels: disk-centre line profiles for the 3D and 1D Holweger--M\"uller models with the O\,\textsc{i} lines artificially shifted to the red. Right panels: limb line profiles for the 3D and 1D Holweger--M\"uller models, using the same shift as in the middle panels. Disk-centre profiles were fitted for abundance, while for the limb profile the disk-centre abundance was used. Lower left panel: synthetic profiles illustrating the effects of the atomic and molecular blends at disk-centre and limb. Shaded regions indicate range over which the disk-centre profiles were fitted. }
  \label{fig:615_profs}
\end{figure*}

The O\,\textsc{i} 615.81~nm line is a triplet, with components at 615.8149, 615.8172 and 615.8187~nm, according to the NIST database. The latter is the strongest component. The components are so close that they are unresolved in the solar spectrum. Its position in the solar spectrum is far from optimal. The line is close to a strong ($\approx 4.8$ pm) \mbox{Fe\,\textsc{i}} line at \mbox{615.77~nm}  and there are multiple weak blends around. The VALD database lists several atomic lines in this region, but besides the strong Fe\,\textsc{i} line only the Nd\,\textsc{ii} 615.782~nm, Ni\,\textsc{i} 615.800~nm and Fe\,\textsc{i} 615.803~nm lines have any measurable effect, though still just barely noticeable. Spectral synthesis using only these lines is still not enough to explain the solar spectrum. However, there seem to be  several molecular lines in this region. These are mostly CN lines, with a few C$_\mathrm{2}$ lines. The molecular lines are important for our analysis and their effect is discussed below. It should be noted that even accounting for the molecules there are still unidentified features in this spectral region, most notably the feature at around 615.793~nm.

The presence of these blends depresses the continuum level around the O\,\textsc{i} lines, so that it becomes a significant source of uncertainty. Our analysis, either in computing the equivalent widths from the observations or fitting the line profiles, suffers from the same problems. For example, if one assumed this oxygen line to have no blends and took the continuum level of the FTS disk-centre intensity atlas, then the equivalent width could be up to $33\%$ higher than our estimate for the oxygen-only contribution. This effect is amplified as one looks at spectra closer to the limb because the different blends vary differently in strength over the solar disk, which makes a precise determination of the continuum level difficult. Unlike the stronger O\,\textsc{i} 777~nm triplet the predicted NLTE effect is rather small for the O\,\textsc{i} 615.8 nm lines: \citet{Asplund2004} predict it to be $-0.03$ dex when neglecting H\,\textsc{i} collisions.

\subsubsection{Wavelength calibration}

A similar approach to the wavelength calibration of [O\,\textsc{i}] 630.03~nm was used. Because a precise wavelength calibration for this line is not as crucial as [O\,\textsc{i}], we derive the wavelength calibration using only the nearby Fe\,\textsc{i} 615.77~nm line. It is important to make the wavelengths of the O\,\textsc{i} consistent with this Fe\,\textsc{i} line because they lie on its red wing, and a different wavelength difference between them would imply a lower or higher blend influence.   %

Using the reference laboratory wavelengths from the NIST database for the Fe\,\textsc{i} 615.77~nm and O\,\textsc{i} 615.81~nm lines and calibrating using the waveshift of Fe\,\textsc{i} lines, the synthetic O\,\textsc{i} profiles are shifted from the observations. This can be seen in the left panels of Fig.~\ref{fig:615_profs}. The 3D synthetic profile is shifted to the blue in regards to the observations, while for the 1D models the opposite happens. None of them matches the observations perfectly. For the 3D case, the shift to the observations is about $0.89\: \kms$, or 1.8 pm. This could mean that there is a significant error in the laboratory wavelengths; the predicted line shifts from the 3D models are inconsistent for the Fe\,\textsc{i} and O\,\textsc{i} lines; or there are blending features causing a shift in the profile. Including known molecular and weak atomic lines is not the explanation. In the top middle panel of Fig.~\ref{fig:615_profs} we have artificially shifted the O\,\textsc{i} lines, so that the synthetic profile matches the FTS intensity atlas. For the rest of the analysis we will be using these artificially shifted profiles (both for 3D and 1D models), but the derived abundance from fitting the disk-centre profiles remains essentially the same if the profiles are not shifted (but as seen in Fig.~\ref{fig:615_profs}, the agreement with the observations is worse). The results for the equivalent widths are not affected by this shift.

\subsubsection{Molecular blends and comparison with observations}

We have included several CN and C$_\mathrm{2}$ lines in the 615.8~nm region (J. Sauval, 2008, private communication). 
The line data are given in Table~\ref{table:615mols}. These data were obtained from from J. Sauval (2008, private communication). The molecular line data is uncertain, especially wavelengths for CN lines. With this uncertainty in mind, we have adjusted some of the wavelengths and $\log gf$ of the CN lines from Table~\ref{table:615mols}. The two (9, 4) band lines at 615.8522 and 615.8796~nm  have been red shifted by 7.2 and 11.6 pm, respectively, so that they match two features in the solar disk-centre spectrum. Furthermore, the $\log gf$ values of these two lines have been increased so that they fit reasonably the observed features. For consistency the $\log gf$ values of the other three CN lines have been increased by the same amount. For the 3D model this increase in $\log gf$ amounts to 0.35 dex; we note that CN lines from different bands do not necessarily require the same correction in abundance or $\log gf$. However, we used the same $\log gf$ correction because the data for the different CN lines have similar uncertainties. Additionally, the (10, 4) band line at 615.8223~nm has also been red shifted by 9.3 pm. 
As detailed below, the inclusion of molecules is of paramount importance to understand the solar spectrum at the limb. 

\begin{figure} 
  \begin{center}
    \includegraphics[width=0.45\textwidth]{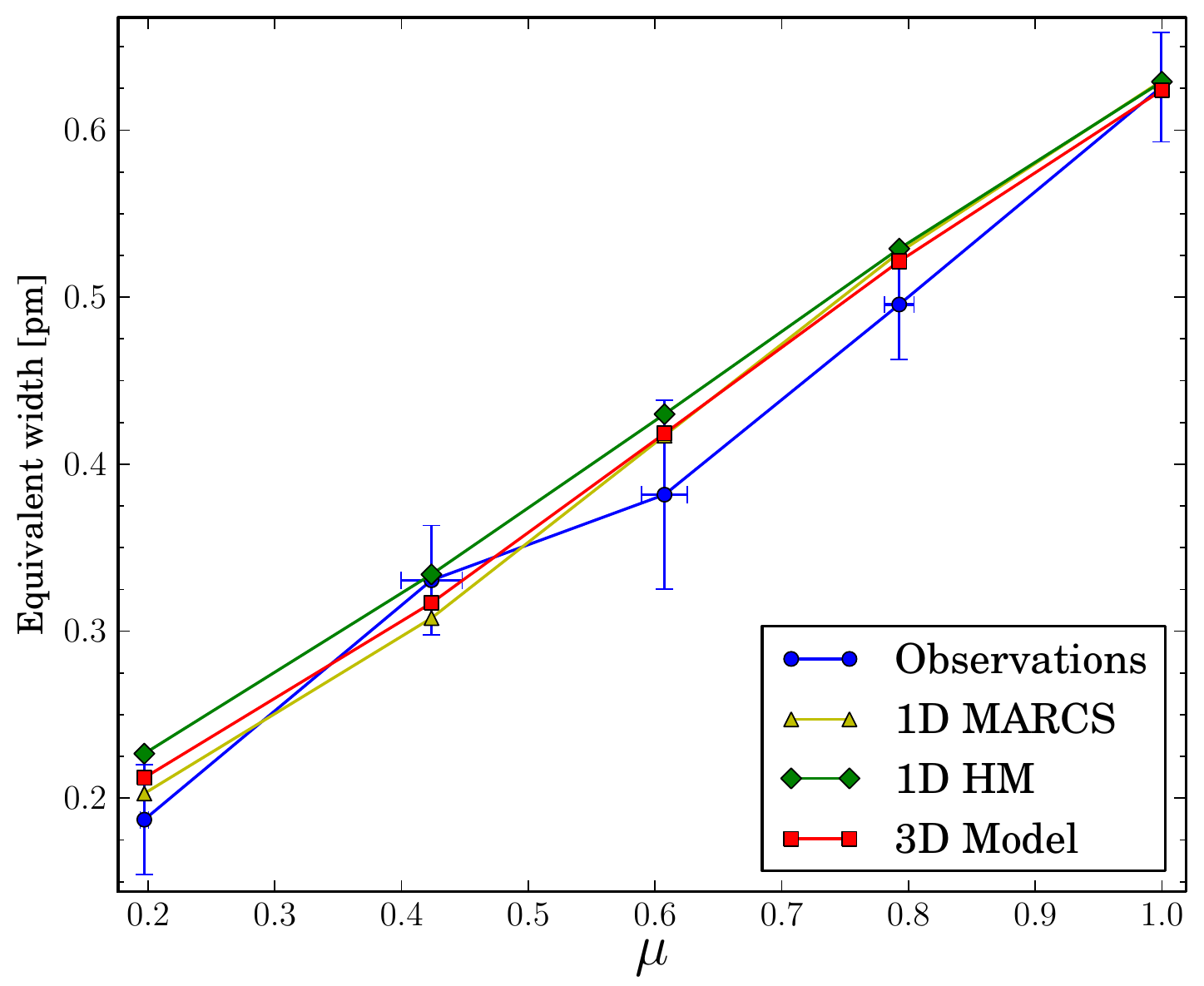}
  \end{center}
\caption{Equivalent width vs. $\mu$ for the three O\,\textsc{i} 615.81~nm lines, including the Fe\,\textsc{i}, CN and C$_\mathrm{2}$ blends.}
  \label{fig:615_eqw}
\end{figure}

Our synthetic profiles included a total of ten blends: the three O\,\textsc{i} lines, the strong Fe\,\textsc{i} line and the molecular lines detailed above. There are other atomic lines in the region, not included because they are very weak.
The profile fits in Fig.~\ref{fig:615_profs} include all these blends. In the lower left panel of Fig.~\ref{fig:615_profs} is shown the synthetic spectrum of only O\,\textsc{i} and Fe\,\textsc{i} compared with the molecules, at the disk-centre and limb. At 10.741 eV, the oxygen lines have a high excitation energy, making them very weak at the higher layers of the atmosphere probed by the limb spectra, when compared to disk-centre. The molecules, on the other hand are formed in cooler regions and are thus significantly stronger at the limb than at disk-centre. As is visible in the figure, at disk-centre the molecules blended with oxygen have a very small contribution to the total line strength, yet at the limb most of the line strength comes from molecules. This type of blending can explain the observations well. As seen in the top right panel of Fig.~\ref{fig:615_profs}, the line profile shape and strength using the oxygen abundance fitted at disk-centre is very close to the observed profile\footnote{The oxygen abundance was the only adjusted parameter in the disk-centre fit, and all other lines have the same abundance and parameters throughout the analysis.}. Because the molecular lines to the blue of the oxygen line dominate at the limb, the line centre is significantly blueshifted as is also confirmed by the observations (the wavelength calibration using the strong Fe\,\textsc{i} line proved important to establish this). Furthermore, the variation in equivalent width in Fig.~\ref{fig:615_eqw} is another confirmation that models including the molecules reproduce the observations satisfactory.

\begin{table} 
\caption{Molecular lines included in the 615.8~nm region.}
\label{table:615mols} 
\centering 
\begin{tabular}{c c c c c c} 
\hline 
Mol. &  Wavelength & System & Band &  $\log gf$ & $E_\mathrm{low}$\\
     &      [nm]   &        &      &            & [eV] \\
\hline
CN             & 615.7850  & A--X  & (7, 2) & -1.498 & 1.800\\ 
C$_\mathrm{2}$  & 615.8106  & $d^3\Pi_g-a^3\Pi_u$  & (0, 2) & -1.652 & 0.508 \\ 
CN             & 615.8109  & A--X  & (6, 1)  & -2.140 & 1.093 \\ 
CN             & 615.8223  & A--X  & (10, 4) & -1.613 & 2.066 \\ 
CN             & 615.8522  & A--X  & (9, 4)  & -1.959 & 1.062 \\ 
CN             & 615.8796  & A--X  & (9, 4)  & -1.924 & 1.088 \\ 
\hline
\end{tabular} 
\end{table} 

The fitted oxygen abundances with the disk-centre FTS atlas were: $\log\epsilon_{\rm O}=8.62$ (3D Model), 8.72 (1D HM), 8.61 (1D \textsc{marcs}). For our own observations at disk-centre the abundance values were higher by about 0.05 dex, difference which can be attributed to the difficulty in finding the continuum level. The carbon and nitrogen abundances used to compute the molecular abundances were different for the 3D and 1D models, and consistent with the abundances derived from each model from other molecular features. For 1D models the macroturbulence was derived from nearby Fe\,\textsc{i} lines, and a microturbulence of $\xi_{\mathrm{turb}}=1.0\: \kms$ was used.

The molecular blends can help to explain the centre-to-limb variation in this spectral region. As seen in Fig.~\ref{fig:615_eqw}, both 1D and 3D models can explain the variations similarly well (although the fitted abundances at disk-centre vary). However, the uncertainty of the molecular line data and the wavelength shift of oxygen make this agreement less convincing. These uncertainties do not allow us to test reliably the line formation across the solar disk, and the effects of different models. Like for O\,\textsc{i} 777~nm, we also experimented with NLTE effects and different $S_\mathrm{H}$. But for the 615~nm lines the NLTE effects are so small that with all the mentioned uncertainties, it is not possible at this time to extract any reliable conclusions about for example the H\,\textsc{i} collisional efficiency.

\subsection{Lines from other species than O I\label{sec:other}}

\begin{figure*}
\begin{center}
  \includegraphics[width=\textwidth]{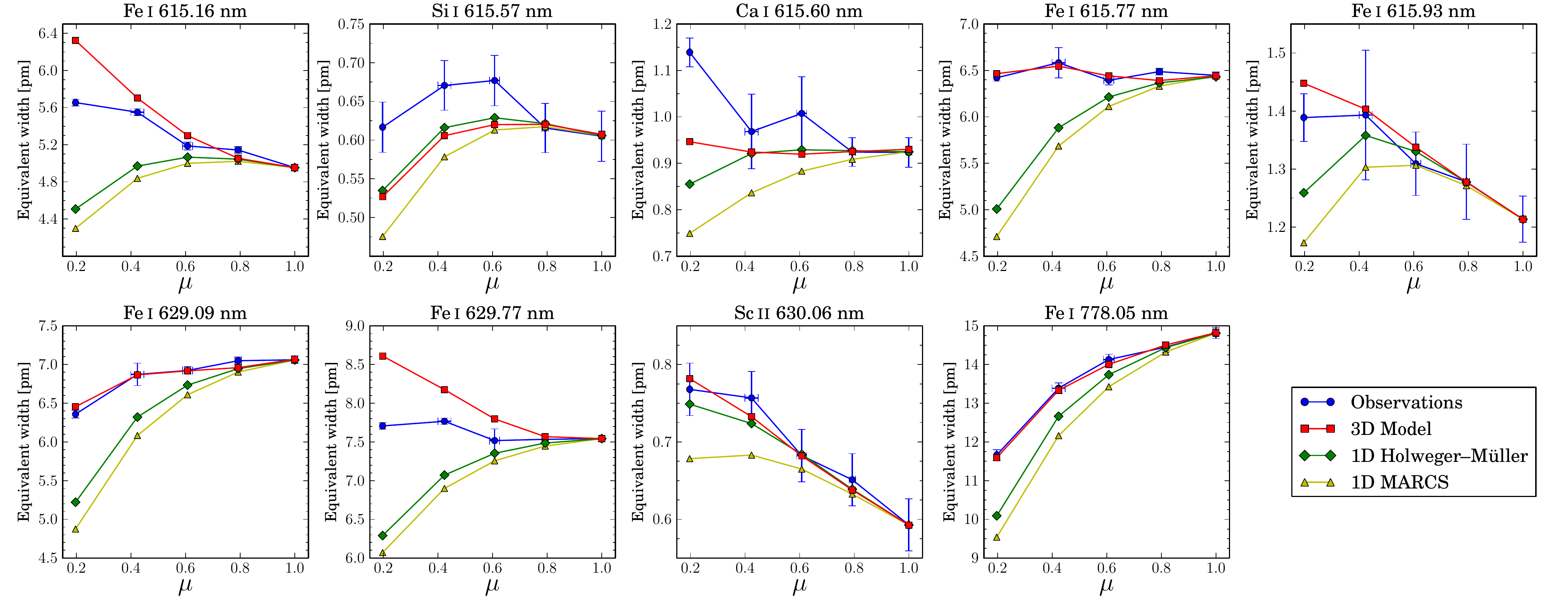}
\end{center}
\caption{Centre-to-limb variations of line equivalent widths. The 1D models were computed with a microturbulence of \mbox{$\xi_{\mathrm{turb}}=1.0\: \kms$} and a macroturbulence extracted from nearby Fe\,\textsc{i} lines. The abundance of each element has been adjusted so that the models fit the observations at disk-centre. The line profiles of Si\,\textsc{i} 615.57~nm and Ca\,\textsc{i} 615.60~nm include weak nearby blends (see text). All the line profiles were computed assuming LTE.}
  \label{fig:meqw}
\end{figure*}

In Fig.~\ref{fig:meqw} we present the centre-to-limb variation of the equivalent width for a few non-oxygen lines included in our observations. The properties of these lines are listed in Table~\ref{tab:lines}. Nearby blends have been included for the Si\,\textsc{i} 615.57~nm and Ca\,\textsc{i} 616.60~nm. Their wings influence each other, so both of them have been included in each of their line profiles, as well as a nearby strong Si\,\textsc{i} line at 615.51~nm. For the Ca\,\textsc{i} 616.60~nm we also include the very weak O\,\textsc{i} triplet at 615.60~nm, which influences the line slightly. Data for all the extra blending lines were extracted from the VALD database.

While the agreement between the 3D model and the lines in Fig.~\ref{fig:meqw} is generally worse than for the oxygen lines it is in most cases an improvement over the 1D models. For most of the Fe\,\textsc{i} lines the 1D models predict a much weaker line at the limb, while the 3D model prediction usually follows the observations. What happens with the Sc\,\textsc{ii} 630.06~nm is very similar with the figure for [O\,\textsc{i}] 630.03~nm. This is probably because this line of a ionized rare element is not very sensitive to the local temperature. In the case of Si\,\textsc{i} 615.57~nm, blends are likely to influence the behaviour of the line. This line is in a relatively crowded spectral region. For Fe\,\textsc{i} 615.93~nm from $\mu\approx 0.4$ to the limb the 1D models predict a decrease line strength, while the 3D model predicts an increase. The observations in this case seem to follow neither and maintain a constant line strength. But it should be noted that weak blending features on the wings of Fe\,\textsc{i} 615.93~nm make the continuum level difficult to determine, hence the big uncertainties in equivalent width. The Fe\,\textsc{i} 629.77 and 615.16~nm lines have a similar variation with $\mu$, which is reasonably reproduced by the 3D model while the 1D models show larger discrepancies and the wrong trend with $\mu$.

At this time we refrain from drawing conclusions about NLTE effects in these lines. However, these may be addressed in the future if there is enough atomic data for these atoms.

\section{Conclusions\label{sec:disc}}

The study of oxygen lines over a range of different positions in the solar disk has proven very fruitful. From departures of LTE to constraints in blends and line data, the detailed line profiles for different $\mu$ values give us a wealth of information not possible before. The relevance of these tests is to ascertain if the temperature structure of the models can reasonably explain the formation of these lines and hence be considered reliable to infer the much debated solar oxygen abundance. Although for abundance analysis the FTS atlases, given their superior resolution and S/N ratio, should be the preferred choice, our high-resolution spectra have proven decisive to identify the shapes and variations of lines and blends over a range of formation temperatures.

Using detailed 3D NLTE radiative transfer we have computed the level populations and synthetic profiles for oxygen lines. The O\,\textsc{i} 777~nm triplet is of particular interest. These lines suffer from NLTE effects, the size of which is determined by inelastic collisions with electrons and neutral hydrogen. Due to the lack of analytical or experimental data regarding the efficiency of the neutral hydrogen collisions, one has to treat it almost as a free parameter, which the centre-to-limb variation of the lines helps getting an empirical estimate for. In a similar approach to \citetalias{CAP2004}, we have determined the empirical value for the multiplication factor of the classical recipe for hydrogen collisions, $S_\mathrm{H}$. For the 3D model and the 1D Holwer--M\"uller model the best fit is for $S_\mathrm{H}\approx 0.85$. %
For the 3D model used by \citetalias{CAP2004} we find a smaller value. These differences illustrate the model dependency of the centre-to-limb derivation of $S_\mathrm{H}$. The 1D \textsc{marcs} model does not even reproduce the observed equivalent widths with any value of $S_\mathrm{H}$.

A more decisive test of the 3D and other models for the O\,\textsc{i} 777~nm lines will have to wait until better data for hydrogen collisions is available. Nevertheless, finding an $S_\mathrm{H}$ that describes the observations provides a self-consistent way of determing the extent of the NLTE effects and to extract a reliable value for the oxygen abundance. As seen in Tab.~\ref{table:abund77}, the oxygen abundance varies quickly with the chosen $S_\mathrm{H}$, especially for  $0.3 \lesssim  S_\mathrm{H} \lesssim  3$, making a self-consistent estimate of the hydrogen collisions of paramount importance. The last remaining discrepancy with the empirical $S_\mathrm{H}$ approach is that the best fitting NLTE profiles at disk-centre are slightly narrower and deeper than the observed. It is not clear if this discrepancy can be improved on the model side or if it is a consequence of using the classical \citet{Drawin1968} formul\ae.

The weak 615.81~nm O\,\textsc{i} line also shows departures from LTE, albeit at a much smaller scale than the 777~nm counterparts. Unfortunately the weakness of the NLTE effects and the presence of blends make it very difficult to estimate $S_\mathrm{H}$ from the centre-to-limb variation of this line. When looking at the line profiles, especially as one gets closer to the limb, we find a line blueshift and strength that are inconsistent with the synthetic spectra, when only oxygen is considered. With a high excitation energy, synthetic models indicate that at the limb the oxygen line should be much weaker than observed. However, if one includes a few molecular blends present in the region, the line strength and blue shift effects can be almost perfectly reproduced in the synthetic spectra. At the limb the molecules blended with the oxygen are stronger than the oxygen contribution. Uncertainties in the input data for the molecular lines and continuum placement errors in the observations mean that the precision of the results is not enough to discern between the different $S_\mathrm{H}$ for the weak NLTE effects.%

Blends also play a determinant role in the [O\,\textsc{i}] 630.03~nm line. The unresolvable blend with Ni in this line has been a major source of uncertainty in an otherwise good indicator for the oxygen abundance (formed in LTE, relatively independent of temperature). A good knowledge of the Ni\,\textsc{i} line properties is essential. For that we use the recent measurement of its $\log gf$ value \citep{Johansson2003} and the recent determination of the nickel abundance for the 3D model \citep{Asplund2009}. Establishing the correct contribution of nickel in the blend is not enough to guarantee a correct determination of the oxygen abundance. Because the oxygen and nickel lines are so close together, a shift in their wavelengths will affect the inferred oxygen abundance, because the fitting procedure will compensate the quantity of oxygen that best matches the observations. It should be noted that because the nickel abundance is kept fixed, the effect of the wavelength shift on the oxygen abundance is not as significant as if it was allowed to vary (\emph{e.g.}, \citetalias{CAP2001}): we estimate it to be less than 0.02 dex. In any case, for this line allowing a wavelength shift in the fitting is not desirable, and a precise absolute wavelength calibration must be obtained. \citet{Ayres2008} made use of Fe\,\textsc{i} lines to calibrate the wavelength scale, as the precision of the FTS atlas by itself is not sufficient. In this work we have used a similar approach of calibration using Fe\,\textsc{i} lines to ensure the accuracy of the results. The line profile fits for the FTS intensity atlas using the 3D model show a remarkable agreement between model and observations, with less than 0.2\% difference for most of the line profile. Being not very sensitive to temperature, this line displays a small centre-to-limb variation in strength. Even with the observational uncertainties inherent of such a weak line, we can see that the 3D model provides a very good description of the observations. This agreement gives us confidence in the 3D model, confirming its robustness to infer the oxygen abundance from this line.

Overall, we find that the 3D model reproduces the observations very well. Despite its one-dimensional nature not allowing for the most accurate shape of the observed profiles, the Holweger--M\"uller model also reproduces the observed equivalent widths, which is not surprising because it is semi-empirical. The 1D theoretical \textsc{marcs} model falls short of its description of centre-to-limb variation of the O\,\textsc{i} 777~nm and [O\,\textsc{i}] 630.03~nm lines, probably because its temperature structure departs from the solar temperature structure.

Using the best fitting $S_\mathrm{H}=0.85$ and abundance results from disk-centre line profile fits we infer an oxygen abundance of 8.68 for the 3D model using the 777~nm lines. This value can be compared the value of 8.66 derived from the [O\,\textsc{i}] 630.03~nm line and the value of 8.62 derived from O\,\textsc{i} 615.82~nm line (the latter dependent on the assumed strength of the molecular lines). The good performance of the 3D model in reproducing not only the observed line shapes and shifts, but also the centre-to-limb variation of the oxygen lines is a good indicator of the quality of the inferred abundances.

\begin{acknowledgements} 
We would like to thank N. Grevesse and J. Sauval for fruitful discussions and for the use of the molecular line database. %
TMDP acknowledges financial support from Funda\c c\~ao para a Ci\^encia e Tecnologia (reference number SFRH/BD/21888/2005) and from the USO-SP International Graduate School for Solar Physics under a Marie Curie Early Stage Training Fellowship (project MEST-CT-2005-020395) from the European Commission. This research has been partly funded by a grant from the Australian Research Council (DP0558836). The Swedish 1-m Solar Telescope is operated on the island of La Palma by the Institute for Solar Physics of the Royal Swedish Academy of Sciences in the Spanish Observatorio del Roque de los Muchachos of the Instituto de Astrofísica de Canarias. %
\end{acknowledgements}

\bibliography{3D}
\bibliographystyle{aa}

\end{document}